\documentclass[fleqn,usenatbib]{mnras}

\usepackage{newtxtext,newtxmath}
\usepackage{fix-cm}

\usepackage[T1]{fontenc}

\DeclareRobustCommand{\VAN}[3]{#2}
\let\VANthebibliography\thebibliography
\def\thebibliography{\DeclareRobustCommand{\VAN}[3]{##3}\VANthebibliography}

\usepackage{graphicx}	
\usepackage{amsmath}	
\usepackage{soul}
\usepackage{multirow}

\title[Short title, max. 45 characters]{Sensitivity of Dry Lava Planet Atmospheric Emission Spectra to Changes in Lava Compositions}

\author[C. P. A. van Buchem et al.]{
Christiaan P. A. van Buchem,$^{1,2}$\thanks{E-mail: vbuchem@strw.leidenuniv.nl (CPAVB}
Rojita Buddhacharya,$^{1,3}$
Mantas Zilinskas,$^{4}$
Sebastian Zieba,$^{5}$ \newauthor
Yamila Miguel$^{1,4}$
and Wim van Westrenen$^{2}$
\\
$^{1}$Leiden Observatory, Leiden University, P.O. Box 9513, 2300 RA Leiden, The Netherlands\\
$^{2}$Department of Earth Sciences, Vrije Universiteit Amsterdam, De Boelelaan 1100, 1081 HZ Amsterdam, The Netherlands\\ 
$^{3}$Astrophysics Research Institute, Liverpool John Moores University, 146 Brownlow Hill, Liverpool L3 5RF, United Kingdom \\
$^{4}$ SRON Netherlands Institute for Space Research, Niels Bohrweg 4, 2333 CA Leiden, The Netherlands \\
$^{5}$ Center for Astrophysics, Harvard \& Smithsonian, 60 Garden St, Cambridge, MA 02138, USA \\
}

\date{Accepted XXX. Received YYY; in original form ZZZ}

\pubyear{\the\year{}}

\begin{document}
\label{firstpage}
\pagerange{\pageref{firstpage}--\pageref{lastpage}}
\maketitle

\begin{abstract}
The atmospheres of hot rocky exoplanets are among the first primary targets of the JWST. Interpreting their atmospheric spectra requires understanding the link between silicate lava compositions and overlying atmospheres. We investigate the sensitivity of simulated emission spectra of dry lava planets to variations in oxide abundances in silicate melt. Our goal is to determine which molten surface features could be distinguishable with future observations. We combine our vaporisation code with gas chemical equilibrium and radiative transfer codes to self-consistently compute atmospheric chemistry and thermal structure. Alongside varying lava compositions, we assess the impact of host star spectral type on emission spectra. TiO$_2$ melt abundance dictates atmospheric TiO, which strongly influences surface temperature and emission spectra due to its short-wave opacity. This creates a degeneracy with heat redistribution efficiency, potentially broken by observing the optical TiO emission feature. Atmospheric SiO and SiO$_2$ abundances depend on melt SiO$_2$ content, with stronger SiO and SiO$_2$ emission features at higher melt abundances. For the currently best observable HREs, changes in TiO$_2$ and SiO$_2$ abundance of about an order of magnitude with respect to BSE, could potentially be observable with 12 JWST eclipse observations. 
\end{abstract}

\begin{keywords}
planets and satellites: atmospheres -- interiors -- composition
\end{keywords}



\section{Introduction}\label{sec:intro}
Hot rocky exoplanets (HREs) serve as potential windows into rocky planet compositions and interiors. Their highly irradiated molten surfaces partially vaporise, significantly influencing the thickness and composition of the overlying atmosphere, providing a direct connection between their interior and their atmosphere \citep{leger_transiting_2009, henning_highly_2018, boukare_deep_2022}. The composition of the atmosphere should therefore be strongly linked to the composition of the surface melt, an idea that has been the premise of a quickly growing body of theoretical and modelling work over the past years \citep[e.g.][]{schaefer_thermodynamic_2004,miguel_compositions_2011,ito_theoretical_2015,kite_atmosphere-interior_2016,nguyen_modelling_2020,zilinskas_observability_2022,wolf_vaporock_2023,van_buchem_lavatmos_2023,seidler_impact_2024}. 

JWST has brought us from theory and modelling into an era where we can start inferring atmospheric compositions of rocky planets \citep{zieba_no_2023,hu_secondary_2024}. With many new observations on the way and the prospect on the horizon of additional measurements by the ELT and Ariel, it is key to understand how the compositions of lava oceans influence the atmospheric properties of HREs and the impact this has on the emission spectra we may observe.

To date, most previous work has assumed that the composition of day-side lava oceans is either equal to that of the bulk silicate Earth (BSE) \citep{palme_cosmochemical_2003}, or to a range of silicate melt compositions typically erupted on the surface of the Earth \citep[e.g.][]{hans_wedepohl_composition_1995,schaefer_chemistry_2009,miguel_compositions_2011,gale_mean_2013, zilinskas_observability_2022}. 

Although this is certainly justifiable as a starting point, it is becoming increasingly clear that the compositional variability of rocky planetary crusts and mantles can be significantly larger than the range covered by these initial assumptions. Estimating the bulk composition of rocky exoplanets \citep{guimond_stars_2024} is an active area of research with estimates being made based on their measured density \citep{rogers_framework_2010,swift_massradius_2011,dorn_can_2015}, the compositions of their host-stars \citep{carter-bond_low_2012,santos_constraining_2017,putirka_composition_2019,putirka_compositional_2021,wang_detailed_2022, seidler_impact_2024} and by analysing polluted white dwarfs \citep{klein_rocky_2011,bonsor_host-star_2021,putirka_polluted_2021,xu_exogeology_2021}. 

In addition, the bulk composition of a rocky exoplanet can only be assumed to be equal to the composition of surface lava if no interior differentiation has taken place - which is unlikely based on observations from our own solar system. Metallic core formation significantly depletes the silicate reservoirs of rocky bodies in iron and siderophile (iron-loving) elements \citep[e.g.][]{wade_core_2005,badro_core_2015,rai_core-mantle_2013,hinkel_starplanet_2018,steenstra_geochemical_2020}. Subsequent crust-mantle differentiation can lead to surface compositions that differ substantially from mantle compositions, as seen for example for silicate-silicate differentiation in \cite{lin_evidence_2017} and graphite-silicate separation in \cite{hakim_thermal_2019,hakim_mineralogy_2019}. 

Temporal changes in melt composition can also take place due to chemical evolution of the melt as a result of the depletion of volatile melt species \citep{schaefer_thermodynamic_2004, kite_atmosphere-interior_2016} or due to crystal formation during lava cooling \citep{villiger_liquid_2004,elkins-tanton_magma_2012,payacan_differentiation_2023}.

Previous work looking at the ways in which melt composition affects HRE emission spectra has been done by \cite{zilinskas_observability_2022} and \cite{seidler_impact_2024}. \cite{zilinskas_observability_2022} limited the number of tested compositions to five. Three of these are based on known compositions from Earth (komatiite, continental crust, and oceanic crust), one on the estimated composition of Mercury, and one is the BSE. The advantage of using "real" compositions is that there is direct evidence for them existing in nature. A large disadvantage is that it limits the variation in the type of melts that one can test. \cite{seidler_impact_2024} used five hypothetical compositions derived from stellar compositions available in the Hypatia catalogue \citep{hinkel_hypatia_2017}. This allows for the testing of a far broader range of different melt compositions, but is still excluding a significant part of the full compositional parameter space.

Due to the uncertainty surrounding the range of actual compositions of a surface lava on a HRE, this work instead investigates how deviations from a canonical BSE model could affect HRE emission spectra. This avoids the issue of having to guess what composition we should expect. Instead, it focuses on understanding how varying a single melt oxide in a BSE composition affects HRE emission spectra, allowing for a better understanding of the importance and influence of the abundance of individual melt species on observable atmospheric features. Our results can be used as a basis from which to interpret observations from JWST and other future efforts such as ELT and Ariel, and to identify, where possible, potential fingerprints in atmospheric emission spectra of specific compositional features of the lava underlying these atmospheres. 

In this work, we assume the HREs to be `dry', by which we mean that we do not include any volatiles species. Although recent work points towards the possibility of HREs supporting volatile-rich atmospheres \citep{hu_secondary_2024,patel_jwst_2024,zilinskas_characterising_2025}, we limit the scope of this study to include non-volatile elements only. 

In the following section, we give a description of how we combined different thermochemical and radiative codes to produce self-consistent pseudo-2D emission spectra for a range of different star-planet system parameters and varying melt compositions. In section \ref{sec:results}, we show the most relevant results out of the wide range of models that we ran, highlighting the potentially distinguishing features in the emission spectra. In section \ref{sec:discussion}, we discuss the caveats and nuances that need to be taken into account when interpreting these results. Finally, we present our conclusions in section \ref{sec:conclusion}.

\section{Methods}\label{sec:methods}

The aim of this paper is to produce emission spectra for volatile-free HREs with a range of different surface melt compositions. In order to achieve this, we took an approach similar to that of \citet{zilinskas_observability_2022}:

\begin{enumerate}
    \item \textbf{Determine the melt composition:} Using BSE \citep{palme_cosmochemical_2003} as a starting point, we vary the abundance of each of the oxides in the melt, keeping the relevant abundances of all other melt components the same. 
    \\
    \item \textbf{Calculate the composition of the melt vapor:} The melt composition and an initial guess of the equilibrium temperature at the chosen point on the planet surface are given as input to our melt-atmosphere chemical equilibrium code LavAtmos \citep{van_buchem_lavatmos_2023}. LavAtmos calculates the equilibrium chemical composition of the vapor coming from the melt. This is then converted to elemental abundances and passed on to the next step. 
    \\
    \item \textbf{Calculate the equilibrium gas chemistry over a grid of temperature-pressure (TP) values:} With the elemental abundances calculated in the previous step and a wide TP grid as input, we use the gas-phase chemical equilibrium code FastChem 3 \citep{kitzmann_fastchem_2024} to calculate the gas chemistry over the entire grid. 
    \\
    \item \textbf{Iteratively determine a TP profile:} Star-planet system parameters and the gas-chemistry over the TP grid from the previous step are given as input to the radiative transfer code \texttt{HELIOS} 3 \citep{malik_helios_2017}. \texttt{HELIOS} produces an atmospheric temperature-pressure profile. 
    \\
    \\
    We then compare the surface temperature found by \texttt{HELIOS} with the initial surface temperature assumed for the melt-vapor calculation done in step \textbf{1}. If these values are not consistent with each other, we repeat step \textbf{1} using the newly calculated surface temperature as input. This loop is repeated until the surface temperature converges. This is generally achieved after 3 to 4 iterations. 
    \\
    \item \textbf{Calculate the equilibrium gas chemistry over the final converged TP profile:} FastChem is used once more to calculate the equilibrium gas-chemistry along the final TP profile. 
    \\
    \item \textbf{Produce the emission spectrum:} We pass on the output of the chemistry and temperature structure to another radiative transfer code petitRADTRANS \citep{molliere_petitradtrans_2019}, which we use to compute the final emission spectrum.
\end{enumerate}

Following these steps allowed us to produce self-consistent 1D models of HRE atmospheres. In the remainder of this section, we explain how we varied the melt composition, the equilibrium chemistry of LavAtmos and FastChem, the radiative transfer codes \texttt{HELIOS} and petitRADTRANS, the pseudo-2D model that we used to produce the final spectra, and finally the different star-planet systems we tested.

\begin{figure*}
    \centering
    \includegraphics[width=1\linewidth]{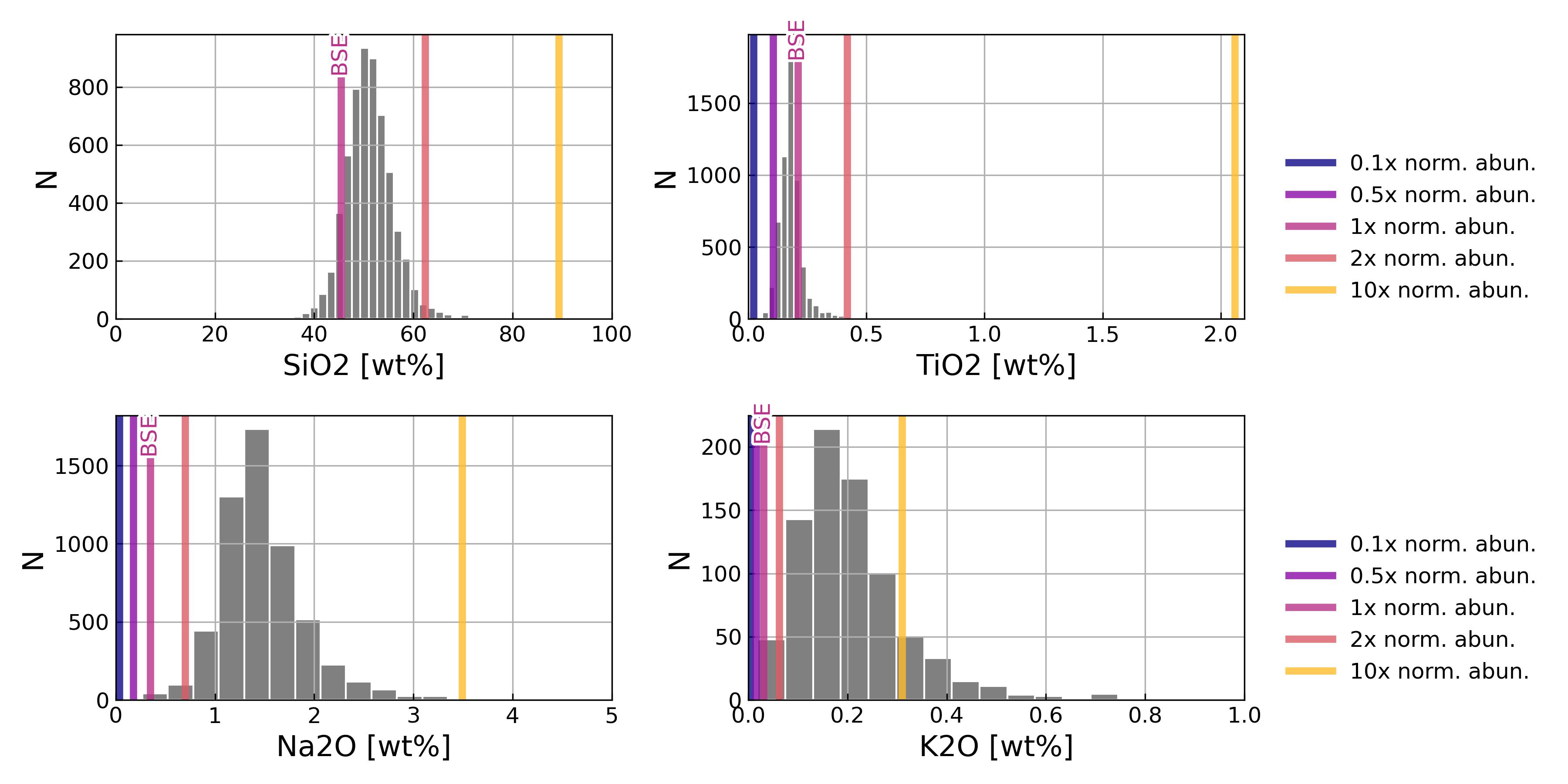}
    \caption{\textbf{Abundance distributions derived from stellar compositions:} Based on the work by \citet{putirka_composition_2019}, we plot the weight percentage distributions of the 4 most impactful melt oxides according to this work. These weight percentages were determined for hypothetical compositions based on the stellar compositions available in the Hypatia catalog \citep{hinkel_hypatia_2017}. The vertical plotted lines represent the abundance ranges in weight percentage that we tested for each oxide, as detailed in the main text.} 
    \label{fig:comp_distribution}
\end{figure*}

\subsection{Melt compositions}\label{subsec:melt_comp}

As explained in section \ref{sec:intro}, significant uncertainty remains about the extent to which lava compositions on the day-side of HREs can differ from a canonical BSE model. Hence, our aim is not to define likely compositions of the lava oceans of HREs but instead to investigate how deviating from the canonical BSE major element composition \citep{palme_cosmochemical_2003} affects atmospheric composition, thermal structure, and emission spectra. 

For each of the eight oxide components included in the canonical melt composition (SiO$_2$, MgO, Al$_2$O$_3$, TiO$_2$, FeO, CaO, Na$_2$O, and K$_2$O) we varied their original BSE abundance individually by factors 0.1, 0.5, 2, and 10 - spanning two orders of magnitude. The only exception was SiO$_2$, which was only modeled for x2 and x10 the BSE abundance due to MELTS (the underlying thermodynamics code used by LavAtmos - see section \ref{subsec:equ_chemistry}) not being able to run for compositions with low SiO$_2$ abundance. 

Within the vaporisation code, given species abundances are normalized to 100 per cent. For example, when increasing the amount of SiO$_2$ in BSE by a factor of 10 from 45.4\% to 454\%, with all other percentages remaining equal to BSE values, the actual bulk composition that is used in the vaporization is renormalized to 89\% with a corresponding decrease of the other oxide species (see appendix \ref{app:melt_composition}). These final normalised values are reflected in the values plotted in Figure \ref{fig:comp_distribution}. In Table \ref{tab:compositions}, we provide an overview of the different weight percentages used for the melt oxide species that we tested as well as their normalized counterparts.  

In Figure \ref{fig:comp_distribution}, we plotted the tested weight percentages for SiO$_2$, TiO$_2$, Na$_2$O, and K$_2$O alongside the distribution of `bulk silicate planet'\footnote{Bulk composition of a planet that excludes an iron-rich metallic core.} compositions derived by \citet{putirka_composition_2019} based on stellar compositions recorded in the Hypatia catalogue \citep{hinkel_hypatia_2017}. We selected these oxides due to them being the most impactful on the emission spectrum of HREs (see section \ref{sec:results}). Figure \ref{fig:comp_distribution} also shows that the tested oxide variations encompass all but a few of the expected edge cases for SiO$_2$ and K$_2$O. The biggest outlier is TiO$_2$, where the 10x abundance is 2.06\% while the maximum value calculated by \citet{putirka_composition_2019} is around 0.42\%. Even so, we consider these abundances to be a conservative estimate of the extremes that could be reached in crust-mantle differentiation and in chemically evolved lava oceans, which we discuss in detail in section \ref{sec:disc_melt_comp}.

\subsection{Equilibrium chemistry} \label{subsec:equ_chemistry}

In order to calculate the composition of the vapor above a melt for a given temperature (step \textbf{1}), we use our thermochemical equilibrium code LavAtmos \citep{van_buchem_lavatmos_2023}\footnote{A recently published update of the code \citep{van_buchem_lavatmos_2025} allows for the inclusion of volatile elements, but we focus the work in this paper on volatile-free atmospheres.}. The partial pressure of any vapor above a melt can be calculated by solving the following equation:
\begin{equation}\label{eq:partial_pressure}
    P_{ij} = K_{r_{ij}}(T,P) a_{j}^{c_{ij}}P_{\text{O}_2}^{d_{ij}}
\end{equation}
Where $P_{ij}$ is the partial pressure of vapor species \textit{i} as formed in the vaporization of melt end-member species $j$, $K_{r_{ij}}$ is the temperature and pressure dependent chemical equilibrium constant of the corresponding vaporization reaction $r$, $a_j$ is the activity of the melt end-member species, $P_{\text{O}_2}$ is the O$_2$ partial pressure, and $c_{ij}$ and $d_{ij}$ are the stoichiometric coefficients necessary to balance the reaction. LavAtmos derives $K_{r_{ij}}$ for each reaction included from the thermodynamic data found in the JANAF database \citep{chase_nist-janaf_1998}. The activity $a_j$ of each end-member species in the melt is calculated by using \texttt{MELTS} \citep{ghiorso_chemical_1995,asimow_algorithmic_1998,ghiorso_pmelts_2002,gualda_rhyolite-melts_2012,ghiorso_h2oco2_2015} through the ThermoEngine python wrapper\footnote{https://gitlab.com/ENKI-portal/ThermoEngine}. Finally, LavAtmos solves for the O$_2$ partial pressure ($P_{\text{O}_2}$) using the laws of mass action and mass balance as constraints. The included oxides are SiO$_2$, MgO, Al$_2$O$_3$, TiO$_2$, FeO, Fe$_2$O$_3$, CaO, Na$_2$O, and K$_2$O.

\citet{zilinskas_observability_2022} relied on published outputs of the \texttt{MAGMA} \citep{fegley_vaporization_1987,schaefer_thermodynamic_2004} code to determine the vapor compositions above a melt. A key difference in the approach taken in this paper is the fact that using LavAtmos allows for far greater flexibility in testing different surface melt compositions. LavAtmos and \texttt{MAGMA} differ from each other mainly in the way in which the activities of the end-member oxide species in the melt are calculated. Instead of \texttt{MELTS}, \texttt{MAGMA} makes use of the Ideal Mixing of Complex Components (IMCC) model \citep{hastie_alkali_1982,hastie_predictive_1985,hastie_predictive_1986}. This leads to some differences in calculated partial pressure (mainly for K); however, they generally agree well with each other, as shown in \citet{van_buchem_lavatmos_2023}. 

Another recently published open-source vaporization code is VapoRock \citep{wolf_vaporock_2023}. This works very similarly to LavAtmos by using \texttt{MELTS} to calculate the activities of liquid oxide species in the melt and solving equation \ref{eq:partial_pressure} for each included vaporization reaction. However, it treats $P_{\text{O}_2}$ as a free parameter given as input by the user instead of constraining it using the laws of mass action and mass balance. A similar approach to this is also taken in the work by \citet{seidler_impact_2024}, who use a revised version of the \texttt{MAGMA} code that allows the $P_{\text{O}_2}$ to be treated as a free parameter. In section \ref{subsec:o2_discussion}, we discuss in detail the pros and cons of using the stoichiometric approach as opposed to the free-parameter approach.

Applying the stoichiometric approach has the additional advantage of isolating the effects of changing melt composition while avoiding the strong effects on atmospheric chemistry caused by large variations of P$_{\text{O}_2}$ (equivalent to fO$_2$ in this context). Over the entire range of different melt compositions tested in this work, the resulting P$_{\text{O}_2}$ values remain between $\Delta\text{IW}$ 1.8 and 1.4, while \cite{seidler_impact_2024} shows that any changes in $\Delta\text{IW}<1$ are likely unobservable. Additionally, the $P_{\text{O}_2}$ calculated by the stoichiometric approach corresponds to the intermediate fO$_2$ case from \cite{seidler_impact_2024} where the spectral features of SiO and TiO are present, but at their weakest. Thus any observed feature is expected to be at least as strong as predicted within the stoichiometric framework and the models presented here show a lower bound on their detectability.

In order to further isolate the effect of changing oxide abundances on the atmospheres and resulting emission spectra of HREs, we do not consider the effect that the presence of volatiles in a melt may have on vaporisation. We assume that no volatile elements (eg. H, C, N, S, and P) are present in either the atmosphere or the melt. 

As explained prior, the chemical composition of the melt vapor is converted to elemental abundances and passed on to the gas chemical equilibrium code FastChem 3 \citep{stock_fastchem_2018,stock_fastchem_2022,kitzmann_fastchem_2024}, which is used to calculate the gas-chemistry of the atmosphere. Initially, this is done over a temperature-pressure grid (step \textbf{2}) and then again over the final TP profile (step \textbf{4}). The majority of the thermal data used by FastChem 3 for this work is from \citet{chase_nist-janaf_1998}, see the appendix of \cite{kitzmann_fastchem_2024} for a full list of sources. Although included in FastChem 3, we have not made use of the condensation due to the fact that the temperatures at all pressure of our models are above 1900 K - which is greater than the point of condensation for all the included gas species \citep{kitzmann_fastchem_2024}. If this work were to be repeated for colder planets or non-inverted atmospheres, then condensation would likely start to play a major role. 

\subsection{Star-planet systems}

\begin{table}
\centering
\caption{\textbf{Overview of tested star-planet systems:} For each spectral type, generic stellar values were used. The orbital distance of each of the planets was determined such that the substellar equilibrium temperature is 3000 K. The planets all have the same assumed radius of 1.5 R$_{earth}$.}
\label{tab:star_types}
\resizebox{\columnwidth}{!}{%
\begin{tabular}{ccccc}
\begin{tabular}[c]{@{}c@{}}Spectral \\ type\end{tabular} & M$_{\text{star}}$ [M$_{\odot}$] & R$_{\text{star}}$ [R$_{\odot}$] & T$_{\text{star}}$ [K] & \begin{tabular}[c]{@{}c@{}}Orbital \\ distance [AU]\end{tabular} \\ \hline
\multicolumn{1}{l}{} & \multicolumn{1}{l}{} &  & \multicolumn{1}{l}{} &  \\
F & 1.3 & 1.23 & 6000 & 2.29e-2 \\
G & 1.0 & 1.00 & 5778 & 1.73e-2 \\
K & 0.8 & 0.84 & 4500 & 8.75e-3 \\
M & 0.5 & 0.57 & 3500 & 3.64e-3 
\end{tabular}%
}
\end{table}

For all of the models, we assume that the planet is $\simeq1.5$ R$_{earth}$. We tested four star-planet systems, each with a star of a different spectral type (F, G, K, and M). For each system, we placed the planet at an orbital distance from the star that corresponds to the distance at which the equilibrium temperature at the substellar point is equal to 3000 K. This was determined using:

\begin{equation}
    d_{\text{orb}} = \text{R}_{\text{star}} \left( \frac{\text{T}_{\text{eq}}}{\text{T}_{\text{star}}}\right)^{2}
\end{equation}

\noindent Where R$_\text{star}$ and T$_\text{star}$ are the radius and temperature of the host star and T$_{\text{eq}}$ is the desired equilibrium temperature. For the G type star, we chose to use the equilibrium temperature and mass of the sun, while for the other three, we used representative temperatures and masses of each spectral type. We used a standard mass-radius relationship of R$_\text{star} \propto$ M$^{0.8}$ \citep{kippenhahn_stellar_2012} to determine the radius of the stars. An overview of the system parameters we used for each spectral type is given in Table \ref{tab:star_types}.

Stellar types hotter than F type star (T$_{star}$ > 7300 K) are not included in this work. This is due to the apparent rarity of HREs orbiting such stars, with only two detections to date \citep{charpinet_compact_2011,morton_false_2016}, and the planet to star flux ratio that is obtained from these systems\footnote{About 2 ppm at 1 $\mu$m and 30 ppm at 20 $\mu$m for a 1.5 R$_{earth}$ planet at an orbital distance of 4.44e-2 AU (equivalent to a substellar equilibrium temperature of 3000K) for a BSE vapor only atmosphere.} is too low to be able to observe accurately with current generation telescopes.

\begin{figure*}
    \centering
    \includegraphics[width=0.9\linewidth]{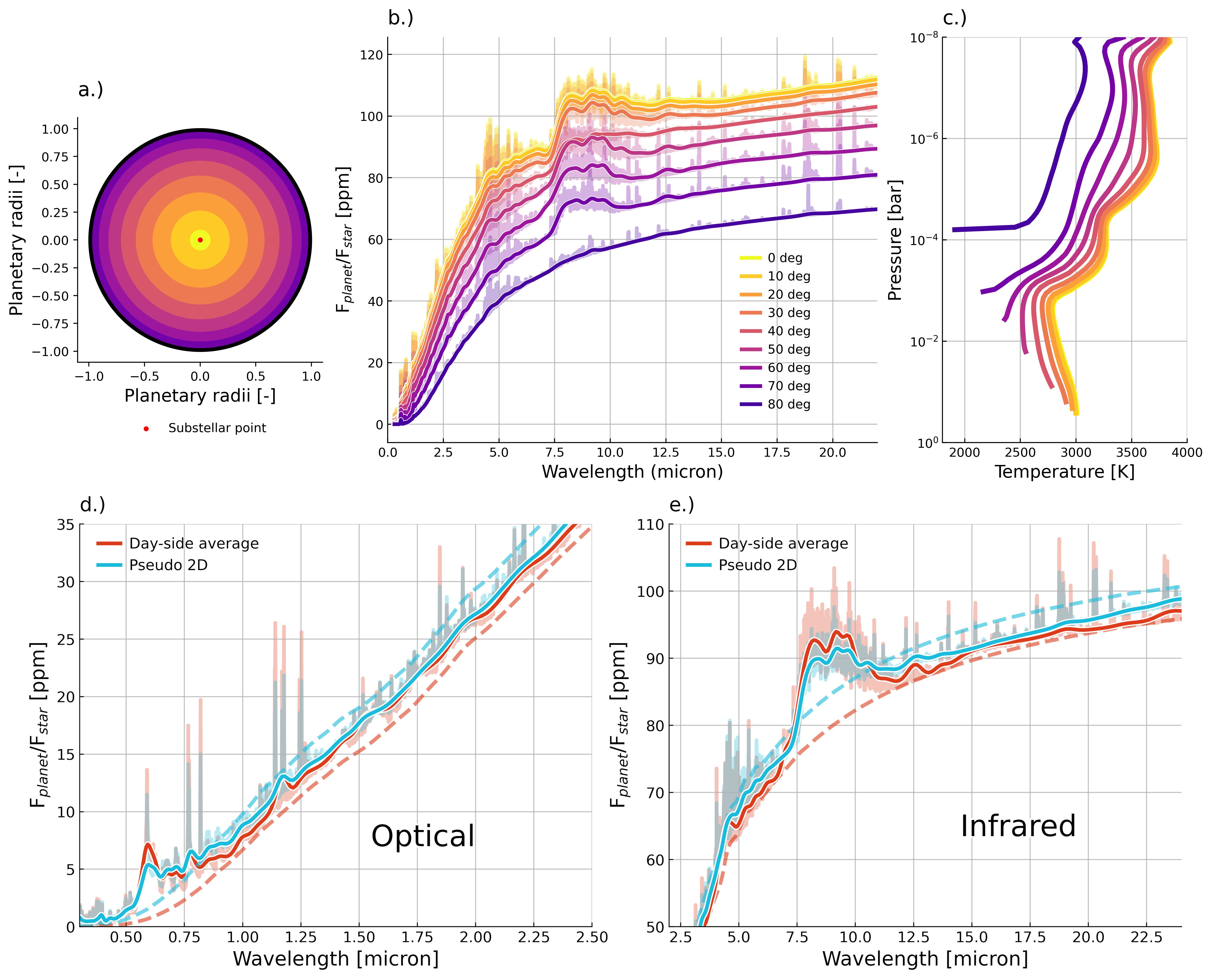}
    \caption{\textbf{Pseudo-2D atmosphere model:} Panel a.) shows the area from which light is emitted from different angles of separation from the substellar point (shown using a red dot). In panel b.) we show the different spectra that are calculated for each of the different rings. Panel c.) shows the TP profiles for each ring, illustrating how the surface temperature (the bottom of the profile) decreases with increasing distance from the substellar point. In panels d.) and e.) we compare the low resolution day-side averaged spectrum (orange) with the pseudo-2D approach (light blue). The higher resolution version of the spectra is are shown using a lower opacity in the background and the spectra for an empty atmosphere (airless rock) on the same planet are shown using the dashed lines. In panel d.) we show the spectra within the optical wavelength range, while in panel e.) we show the spectra within the infrared wavelength range.}
    \label{fig:pseudo2D_model}
\end{figure*}

\subsection{Radiative transfer}\label{sec:rad_trans}

In order to calculate the TP profiles of the tested atmospheres (step \textbf{3}), we used the radiative-transfer code HELIOS \citep{malik_helios_2017,malik_analyzing_2019,whittaker_detectability_2022}. As in \citet{zilinskas_observability_2023}, these TP profiles were self-consistently calculated with the output of FastChem and LavAtmos, such that the final surface temperature is consistent with the overall atmospheric chemistry. As explained in step \textbf{4} at the start of the methods section, this is achieved by iterating until the surface temperature used to calculate the vapor composition is the same as the temperature at the bottom of the TP-profile. 

In all cases, we used the standard diatomic adiabatic coefficient $\kappa = 2/7$ for convective adjustment. For our opacities, we used all of the atomic and molecular species that are included in our melt compositions. This includes atoms Al, Ca, Fe, K, Mg, Na, Si, Ti, as well as molecular species AlO, CaO, MgO, O$_2$, SiO, SiO$_2$, and TiO\footnote{Although FeO is abundant in lava-vapor atmosphere, there are currently no available line lists for this molecule and is therefore not included in the radiative transfer model.}. The line lists were kept the same as in \citep{zilinskas_observability_2023}, with atomic species using the Vienna Atomic Line Database (VALD3) \citep{ryabchikova_major_2015} and the molecular species from a range of sources. All of the line lists that we used and their corresponding sources can be found in Table \ref{tab:opacity_sources}. Note that these opacities were sourced from the DACE\footnote{ https://dace.unige.ch/} database or calculated using HELIOS-K \citep{grimm_helios-k_2015,grimm_helios-k_2021}. In Figure \ref{fig:opacites}, we show the unweighted opacities of some of the most important atmospheric species. 

We used the \texttt{PHOENIX} models \citep{husser_new_2013} to calculate the stellar spectra for each of the F, G and K spectral types. Just as in \citet{zilinskas_observability_2023}, we combined the \texttt{PHOENIX} spectrum for an M type star with scaled shortwave \texttt{MUSCLES} spectra \citep{france_muscles_2016,loyd_muscles_2016,youngblood_muscles_2016} so as to accurately model the increased shortwave emission seen in cooler stars. See Figure \ref{fig:stellar_spectra} for a side by side comparison of the included stellar spectra.  

For step \textbf{6}, we pass the output of the chemistry and temperature structure to \texttt{petitRADTRANS}\footnote{Version 2.7.6} \citep{molliere_petitradtrans_2019,molliere_retrieving_2020}, which we use to compute the final emission spectrum. Where possible, the opacity line lists were kept the same as in \texttt{HELIOS}, but the opacities themselves were obtained from ExoMol\footnote{https://www.exomol.com/} prepackaged correlated-k tables \citep{chubb_exomolop_2021}. All of the spectra were calculated using the low-resolution mode of $\lambda/\Delta\lambda = 1000$, with a wavelength range of 0.3 -- 28 \textmu m.

\subsection{Pseudo-2D model}\label{sec:pseudo_2D}

In order to alleviate some of the drawbacks of modelling thermal emission from a 3D object using a 1D approach, as previously done in \citet{zilinskas_observability_2022,zilinskas_observability_2023}, we expanded our method to a pseudo-2D approach. This was first used in \citet{zieba_k2_2022} to model the atmosphere of the planet K2-141 b and has since then also been used to rule out a melt vapor-only atmosphere for 55-Cnc e in \citet{hu_secondary_2024}. 

In Figure \ref{fig:pseudo2D_model}, we illustrate how this approach works in detail. We calculate the 1D emission spectrum at 10$^{\circ}$ intervals moving away from the substellar point (panel \textit{a} in Figure \ref{fig:pseudo2D_model}). Moving away from the sub-stellar point increases the zenith angle, leading to less stellar irradiation hitting the surface of the planet, hence, lowering the surface temperature. This is demonstrated in panel \textit{c}. The spectra calculated at each angle are then weighted according to the surface area covered by each interval and summed, giving a single averaged spectrum of the entire day-side of the planet. This approach allows us to take into account how melt vapor composition changes as a function of surface temperature and the effect that this has on the spectral features in the emission spectrum.

In Figure \ref{fig:pseudo2D_model}, we also show a comparison between this approach (shown in light blue) and another commonly used approach of using a day-side averaged temperature estimation (shown in orange). The models were made using a bulk silicate earth (BSE) composition \citep{palme_cosmochemical_2003} around a G type star. In this comparison, we can see that the pseudo-2D approach leads to slightly less pronounced features and a slight increase in the continuum spectrum, most notably at wavelengths beyond 5 $\mu$m (panel \textit{b}). Although the differences between these two approaches are relatively minor when considering low-resolution spectra, these may become more important when high-resolution spectra of HREs are analysed.

\section{Results}\label{sec:results}

\begin{figure*}
    \centering
    \includegraphics[width=0.9\linewidth]{figures/contribution_locations_combined.jpg}
    \caption{\textbf{Overview of a vapor atmosphere:} Each of these panels represent different aspects of a vaporised atmosphere above a BSE melt for a G-type star at 40$^\circ$ from the substellar point. In the top panel (\textit{a}) we compare the emission spectrum when including all species (green) with cases where a single specified opacity is excluded. For the no-atmosphere model we assume the planet to be a black body radiating at 3000 K. The panel below that, (\textit{b}), shows the difference in ppm between the spectrum containing all species and the ones with the opacity of one species removed. The bottom left panel (\textit{c}), contains both the temperature-pressure profile of the atmosphere (in black) as well as the VMR of a few keys species in the atmosphere. Finally, in the bottom right panel (\textit{d)}) the contribution to the emission as function of pressure and wavelength is visualised. We highlighted the regions where the contribution from key species is most prominent.}
    \label{fig:contribution_plot}
\end{figure*}

As explained in section \ref{sec:methods}, we produced models for F, G, K, and M type stars, varying the melt oxide abundances of SiO$_2$, MgO, Al$_2$O$_3$, TiO$_2$, FeO, CaO, Na$_2$O, and K$_2$O. We found that the melt species that had the strongest influence on the temperature-pressure profile and emission spectrum of a generic HRE with a substellar temperature of 3000 K are TiO$_2$ and SiO$_2$, supporting some of the findings of \cite{seidler_impact_2024}. This is covered in detail in the following subsections of this paper along with some tentative indications that Na and K could also play a potentially observable role in HRE emission spectra and TP profiles. For the remaining species, however, varying their abundance in the melt oxides did not lead to any significant effects on the shape of the emission spectra and TP profiles and are therefore not shown in the results.

Before delving into the results, we first devote a subsection to analysing in detail the temperature-pressure and chemical structure as well as the opacity contributions and the resulting emission spectrum of a BSE atmosphere for a G-type star with an equilibrium temperature of 3000 K. This will help explain and interpret the results in the following sections.

\subsection{The anatomy of a BSE vapor atmosphere}

To understand how a changing melt and atmosphere composition affects the emission spectrum of a HRE, it is useful to first visualise where the emission originates in the atmosphere. In Figure \ref{fig:contribution_plot}, we show the chemistry, the emission spectrum, and the contribution function of a rock vapor atmosphere above a melt with a BSE composition. For this model, we assumed the same parameters as that of a planet orbiting a G-type star (see Table \ref{tab:star_types}). In this figure, we show a spectrum of a model which assumes incident irradiation equivalent to that at 40$^\circ$ away from the substellar point due to this resulting in similar temperatures as the averaged temperature of a pseudo-2D model (see Section \ref{sec:pseudo_2D})\footnote{We show the structure of the atmosphere for a single incident angle and not for an averaged TP profile due to the TP profiles varying strongly depending on the incident angle.}.

In the top panel (\textit{a}) of Figure \ref{fig:contribution_plot}, we show the calculated emission spectrum for the case with no atmosphere (grey), an atmosphere containing the full list of species (green, see Table \ref{tab:opacity_sources} for details on each species), and three spectra calculated with a single species omitted - TiO, SiO, and SiO$_2$ (blue, red, and maroon respectively). In order to highlight the changes in flux when removing these species from the opacities, we added a panel below (\textit{b}) in which we plot the difference in ppm between the spectra. It should be noted that for the spectra with the omitted species, we only left out the opacity of these species when calculating the emission spectrum using petitRADTRANS (step \textbf{6} in Section \ref{sec:methods}), so the temperature-pressure profile was calculated assuming the presence of all species and remains the same. The leftmost panel (\textit{c}) contains both the temperature-pressure structure of the atmosphere (in black, x-axis located at the bottom), as well as the volume mixing ratios (VMR) of a selection of atmospheric species (colored, x-axis located at the top). The panel to the bottom right (\textit{d}) shows the opacity contribution as a function of wavelength and pressure. Regions with especially strong contributions from SiO, SiO$_2$, and TiO are highlighted (red, maroon, and blue respectively). 

Looking at the spectrum for the case with no atmosphere in panel \textit{a}, we see that including an atmosphere leads to an overall decrease in flux. This is due to the major emitting region (also known as the photosphere) in the model containing an atmosphere being located in a region of the atmosphere (between 1e-1 and 1e-2 bar) which has a lower temperature than the surface. Similarly, the SiO$_2$ features around 7 $\mu$m show up as absorption features in the spectrum due to originating from a relatively colder part of the atmosphere - coinciding with 1e-3 bar, the lowest temperature region of the atmosphere.

As we go beyond this point to higher altitudes and lower pressures around 1e-4 bar, TiO and SiO become the spectrally dominant species. As explained in \citet{gandhi_new_2019}, thermal inversions occur in atmospheres where
\begin{equation}\label{eq:inversion_crit}
    \kappa_{\text{vis}}/\kappa_{\text{ir}} \gtrsim 1
\end{equation}
Where $\kappa_{vis}$ is the average opacity in the optical wavelength range (between about 0.1 and 1 $\mu$m) and $\kappa_{ir}$ the average opacity in the infrared (beyond $\simeq 3 \mu$m). Looking at the contribution plot (panel \textit{d}), we see a strong increase in the opacity in the optical, which, despite the SiO opacity contribution in the infrared, is enough to satisfy the criterion in equation \ref{eq:inversion_crit}, leading to a thermal inversion. The presence of a thermal inversion allows for emission features in the spectrum \citep{gandhi_new_2019}, which we see clearly for TiO around 0.7 $\mu$m and for SiO at 3, 4.5, and 8 micron. Going up further into the atmosphere, the opacity contribution of these species decreases, coinciding with the decrease in their abundance at lower pressures, as can be seen by comparing panel \textit{c} to panel \textit{d}. The more complex species disassociate into their atomic counterparts with increasing temperature and decreasing pressure. 

Thanks to the strong influence on both the thermal structure of the atmosphere and the emission spectrum of SiO, SiO$_2$, and TiO, we find that varying the abundance of SiO$_2$ and TiO$_2$ in the melt has a significant effect on the emission spectra of HREs. This is covered in detail in the following subsections. Besides the aforementioned species, there are a few other species that may have an observable effect on the emission spectrum of HREs \citep{zilinskas_observability_2022}. However, we find that only Na and K could potentially lead to some observable changes if their corresponding melt end-member species (Na$_2$O and K$_2$O) are changed. This is explored in the final subsection of the results.

\begin{figure}
    \centering
    \includegraphics[width=\linewidth]{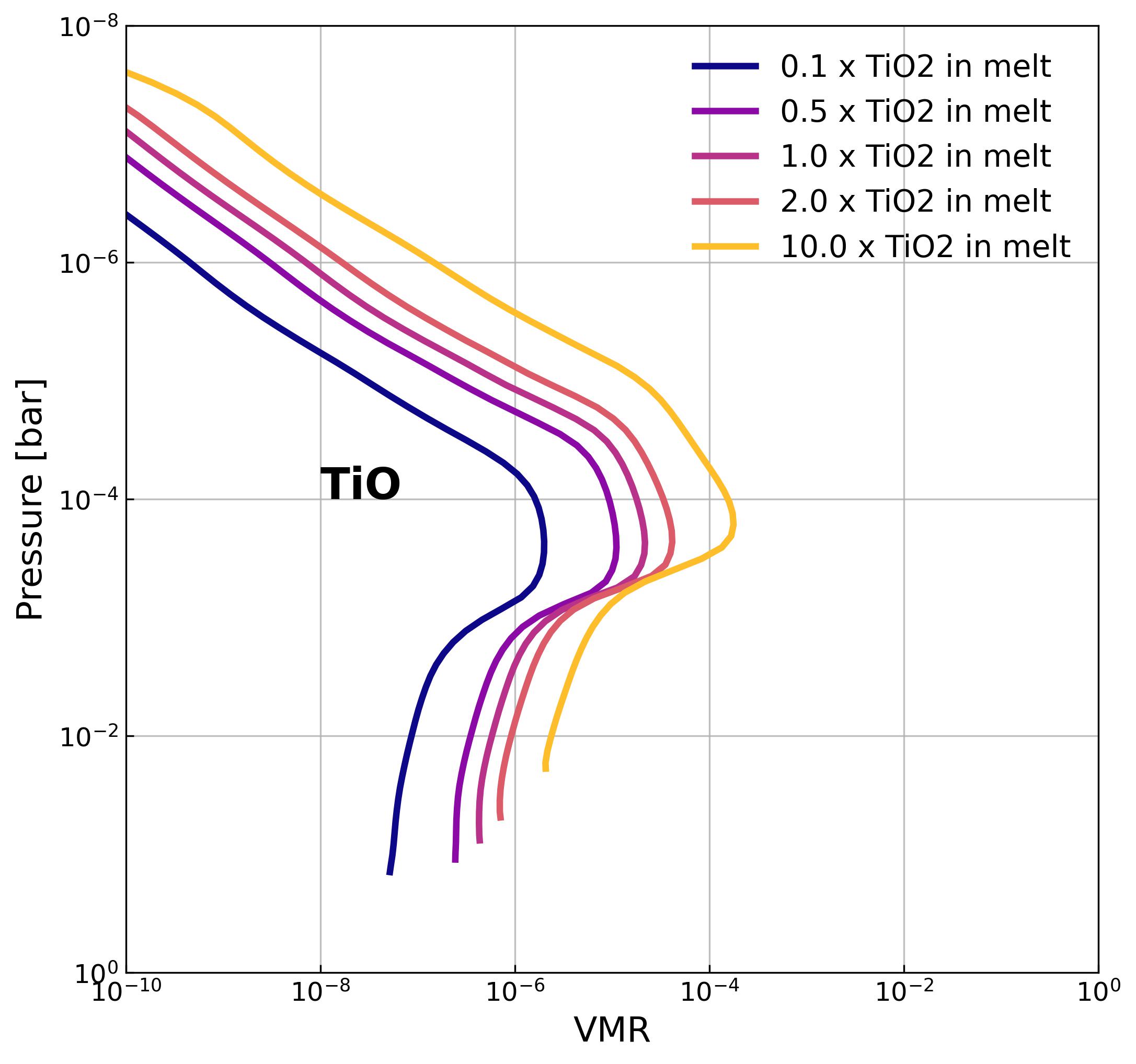}
    \caption{\textbf{Atmospheric TiO changing with TiO$_2$ melt abundance:} Shown for a planet orbiting a G-type star (see Table \ref{tab:star_types} for parameters). The TiO$_2$ abundance in the melt was changed with respect to the BSE abundance (see Table \ref{tab:compositions}).}
    \label{fig:TiO2_VMR}
\end{figure}

\begin{figure*}
    \centering
    \includegraphics[width=0.86\linewidth]{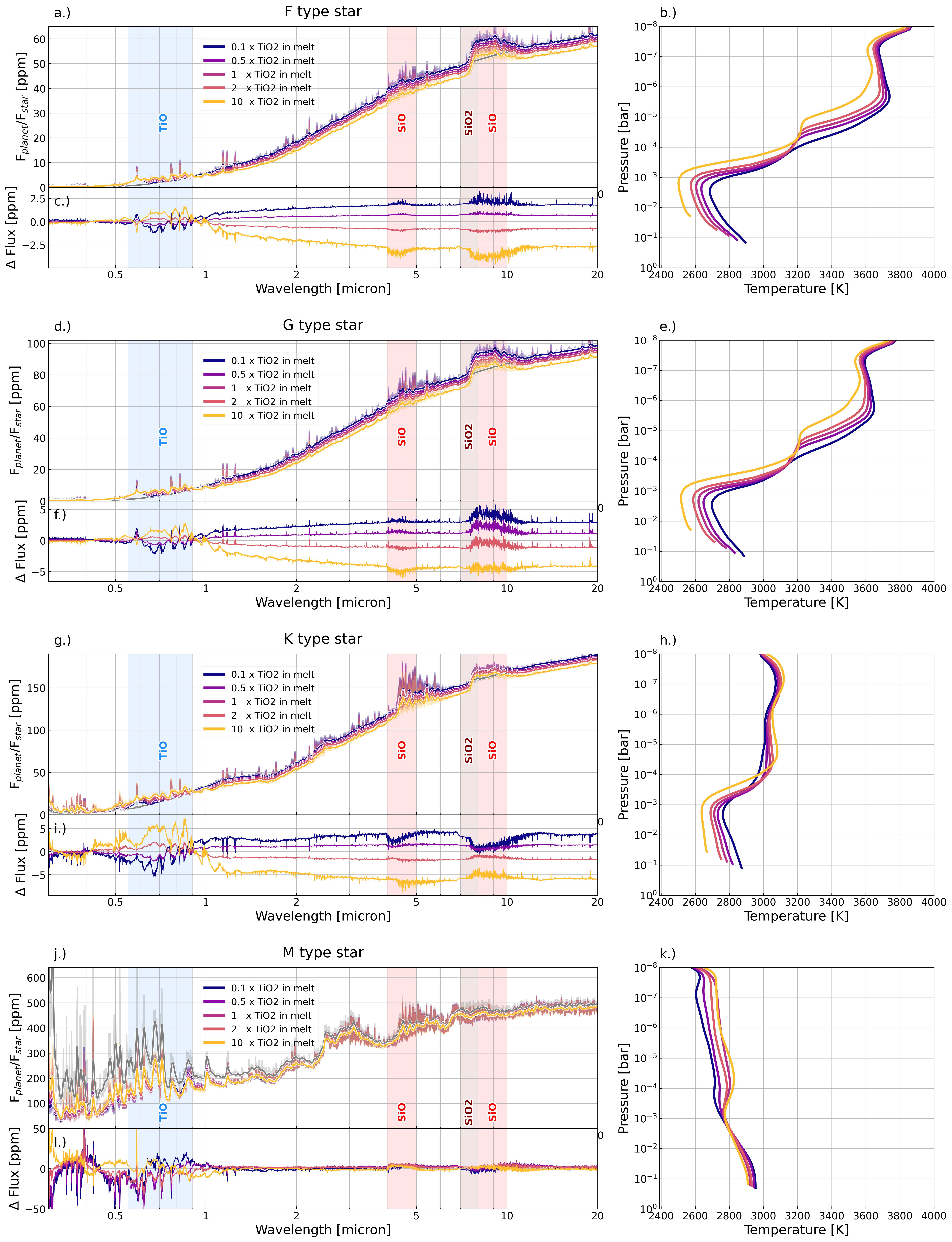}
    \caption{\textbf{Effect of varying TiO$_2$ melt percentage on emission spectra} -  For an F-, G-, K-, and M-type star we show the calculated pseudo-2D emission spectra and TP profiles for different TiO$_2$ melt abundances. Under each spectrum we also plot the differences in flux between the different spectra with respect to 1 x TiO$_2$ melt (which is the same as a BSE composition). In grey we plotted the emission spectra of a model without any atmosphere for reference. The TP profiles shown are those for the models calculated at a 40$^\circ$ angle away from the substellar point. Note the difference in the scales of the y-axes of the emission spectra for the different stellar types.}
    \label{fig:TiO2_results}
\end{figure*}

\subsection{TiO2 variation}

In Figure \ref{fig:TiO2_VMR}, we show how varying the TiO$_2$ melt abundance affects the VRM of TiO throughout the atmosphere for a planet orbiting a G-type star\footnote{The relative behavior of TiO shown in Figure \ref{fig:TiO2_VMR} is representative for the behavior of F-, K- and M- type stars as well.}. The change in VMR of TiO at a given pressure is almost in linear relation to the change in abundance of TiO$_2$ in the melt. However, we do see a decrease in surface pressure as a result of decreasing surface temperature. In Figure \ref{fig:TiO2_results} we show in detail how a varying TiO$_2$ melt abundance affects the emission spectrum (panels in the left column), and the temperature pressure profile (panels in the right column) for F, G, K, and M type stars. The temperature profiles shown are those calculated at a 40$^\circ$ angle away from the sub-stellar point as they were deemed most representative of the average dayside (see Figure \ref{fig:pseudo2D_model}). To allow for a clearer visualization of the differences in the spectra, we added a panel below panels containing the emission spectra showing the difference in flux between the spectra calculated using the BSE (indicated with the `1 x TiO$_2$' label) composition and those with varying levels of TiO$_2$. 

The behavior of the emission spectra and TP profiles is qualitatively similar for planets orbiting F-, G-, and K- type stars, which we will describe first, after which we will focus on the results of planets orbiting an M- type star. Changing the TiO$_2$ melt abundance - and as a result the TiO atmospheric abundance - of these planets has a significant effect on both the temperature-pressure profile of the atmosphere and its emission spectrum. The two most notable effects are 1) a significant decrease in the surface temperature and 2) an increase in the size of the TiO emission features (highlighted in blue in Figure \ref{fig:TiO2_results}). 

The decrease in surface temperature, from around 2900 K to 2550 K for the F- and G-type stars and from 2850 to 2650 K for K-type stars, is due to the increase in TiO abundance leading to an increase in the opacity of the atmosphere in the optical wavelengths (see panel \textit{d} in Figure \ref{fig:contribution_plot}). This allows less of the stars irradiation to reach the planet's surface, decreasing its temperature, leading to a significantly lower flux being emitted from the planet, as is shown in panels \textit{c}, \textit{f}, and \textit{i}, where we see a decrease in flux of up to 5 ppm in the infrared (with respect to the BSE case) for a 10x increase in TiO$_2$ abundance. 

In the upper atmosphere, however, the temperatures converge again up around 3900 K. Hence, as the TiO$_2$ melt abundance (and the TiO atmospheric abundance) increases, the size of the temperature inversion increases as well. Looking back at equation \ref{eq:inversion_crit}, we see that this is what one would expect for an increase in the opacity in the optical wavelength range. The result of the strengthening of the temperature inversion and increasing TiO atmospheric abundance with increasing TiO$_2$ melt abundance, is that even with an overall decrease in flux, the TiO emission features grow. 

Comparing the low TiO$_2$ abundance TP profiles with their high abundance counterparts for F- and G- type stars also shows that in atmospheres with abundant TiO$_2$ the TP profile becomes almost isothermal between 1e-4 and 1e-5 bar - which coincides with the pressures with the highest TiO abundance (see Figure \ref{fig:TiO2_VMR}) and also the pressures from which the greatest opacity contribution in the optical is located (see Figure \ref{fig:TiO2_results}). 

All of the changes to the emission spectrum and temperature pressure profile caused by varying the TiO$_2$ abundance, are due to the changing absorption of irradiation within the optical. Hence, the spectra of planets orbiting F- and G- type stars, which emit the majority of their energy in the optical wavelengths (see Figure \ref{fig:stellar_spectra}), are affected more than for cooler K-type stars. This is most visible in the temperature pressure profiles in Figure \ref{fig:TiO2_results}. Those calculated for K-type stars have far weaker thermal inversions and are close to isothermal above 1e-4 bar. This trend of decreasing thermal inversion strength was also found for hot-Jupiters by \citet{lothringer_influence_2019} and was also shown for HREs in \citet{zilinskas_observability_2022}. What should be noted though, is that the planet to star flux ratio is higher for cooler/smaller stars, hence the differences in the size of the TiO$_2$ emission features are greater than that of hotter stars, reaching up to 5 ppm, despite the smaller temperature inversion. 

Looking at the spectrum of a planet orbiting an M-type star in the bottom row of Figure \ref{fig:TiO2_results}, we see that varying TiO$_2$ melt abundance does not seem to have a very significant effect on the emission spectrum or TP profile of the planet. This is due to the fact that the emission peak of M-type stars is shifted so far to the red (see Figure \ref{fig:stellar_spectra}) that the high opacity of TiO in the optical does not block as significant an amount of the radiation as for the hotter stars. Hence, the surface temperature is not affected as strongly and the overall flux level remains about the same for all TiO$_2$ melt abundances. Furthermore, the relatively small temperature inversion and generally lower temperature of the atmosphere leads to far less prominent emission features, making it more difficult to discern the planetary atmosphere features from the features in the stellar spectrum. 

\begin{figure}
    \centering
    \includegraphics[width=\linewidth]{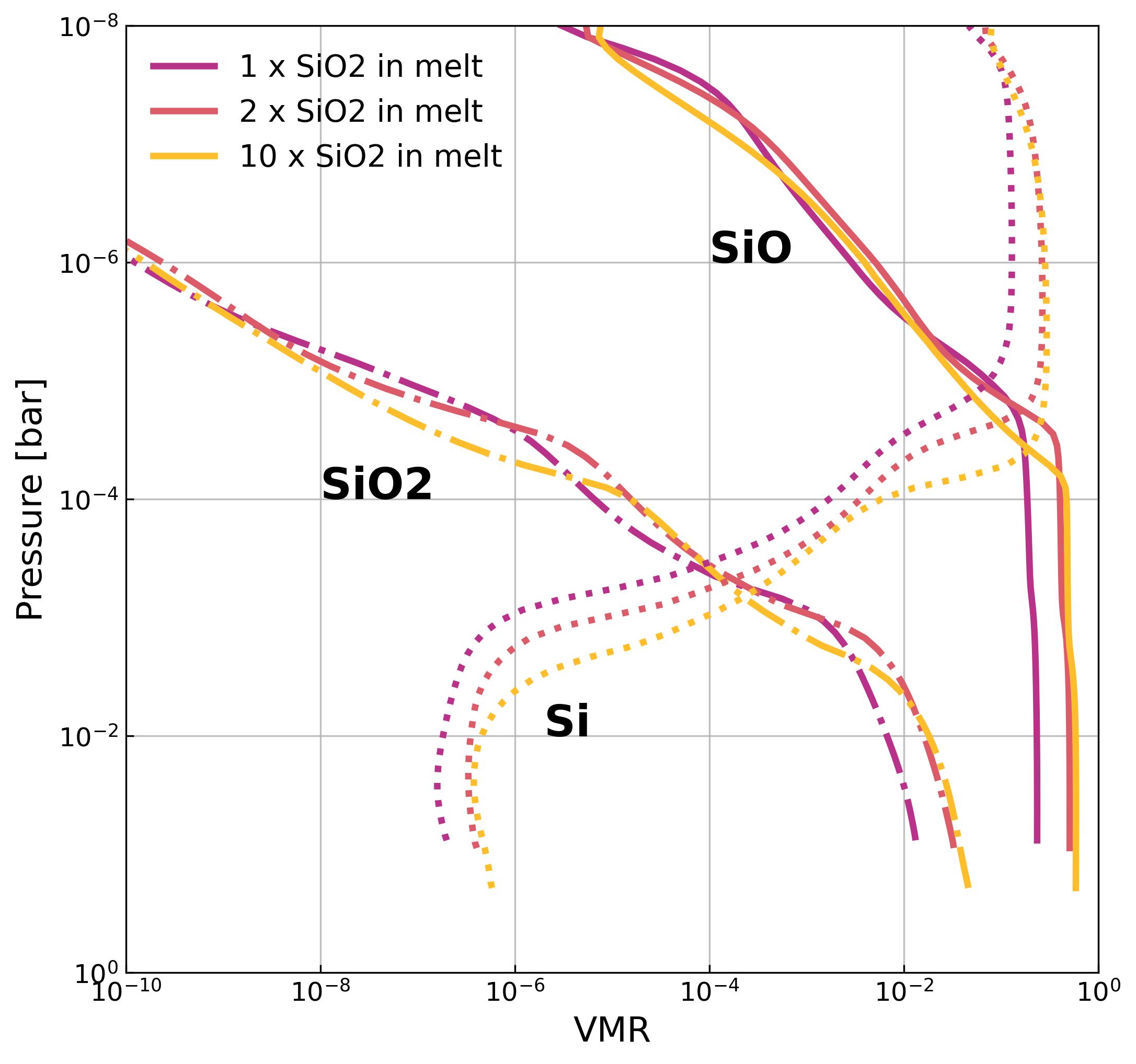}
    \caption{\textbf{Atmospheric SiO and SiO$_2$ changing with SiO$_2$ melt abundance:} Shown for a planet orbiting a G-type star (see Table \ref{tab:star_types} for parameters). The SiO$_2$ abundance in the melt was changed with respect to the BSE abundance (see Table \ref{tab:compositions}).}
    \label{fig:SiO2_VMR}
\end{figure}

\begin{figure*}
    \centering
    \includegraphics[width=0.85\linewidth]{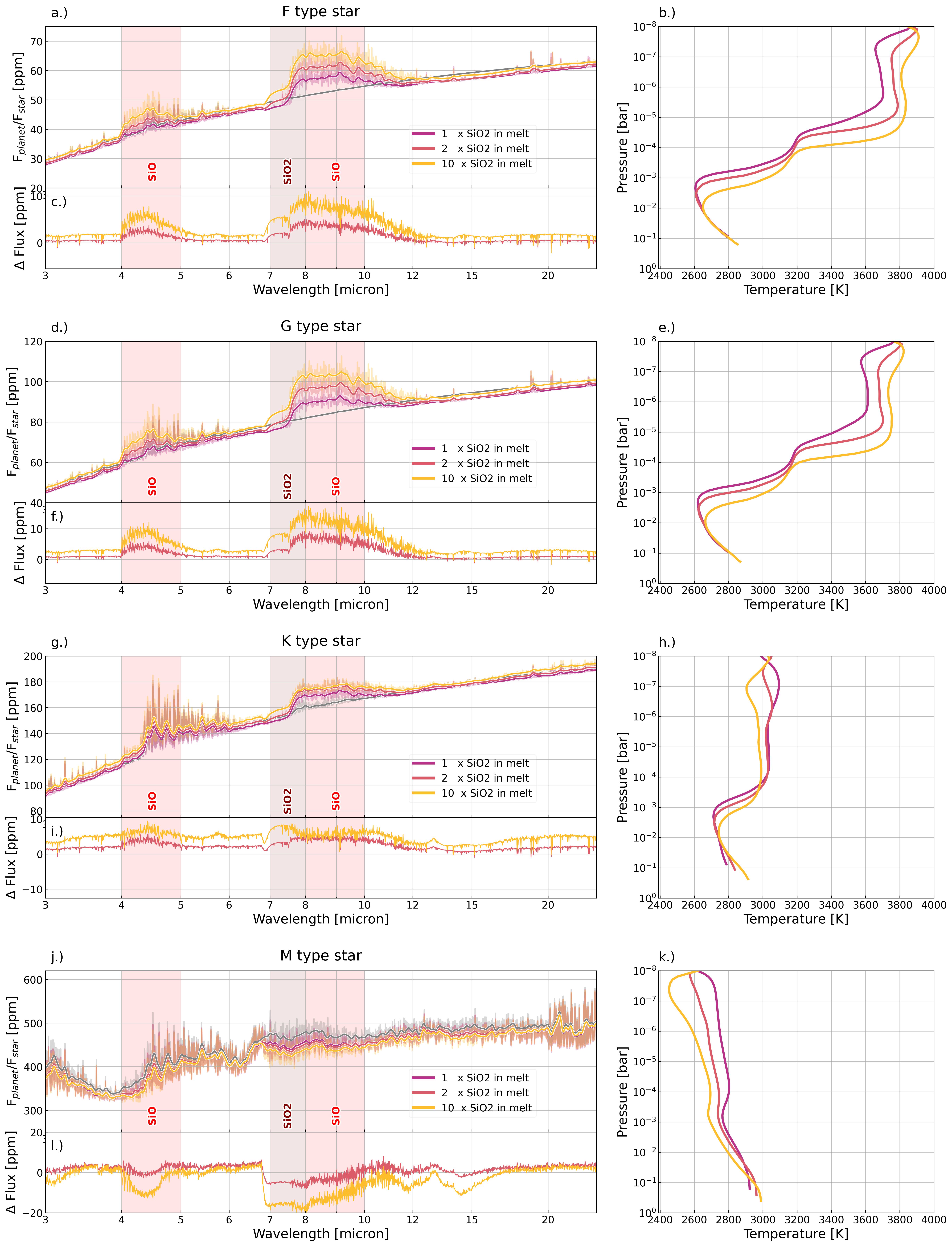}
    \caption{\textbf{Effect of varying SiO$_2$ melt percentage on emission spectra} -  For an F-, G-, K-, and M-type star we show the calculated pseudo-2D emission spectra and TP profiles for different SiO$_2$ melt abundances. Under each spectrum we also plot the differences in flux between the different spectra with respect to 1 x SiO$_2$ melt (which is the same as a BSE composition). In grey we plotted the emission spectra of a model without any atmosphere for reference. The TP profiles shown are those for the models calculated at a 40$^\circ$ angle away from the substellar point. Note the difference in the scales of the y-axes of the emission spectra for the different stellar types.}
    \label{fig:SiO2_results}
\end{figure*}

\subsection{SiO2 variation}

As mentioned in section \ref{sec:methods}, we were not able to model a decrease in SiO$_2$ abundance in the melt composition due to \texttt{MELTS} not being able to run for compositions with low SiO$_2$ abundance. In Figure \ref{fig:SiO2_VMR} we show the VMR of the most important Si species in the atmosphere of a G-type star at 40$^\circ$ degrees away from the substellar point. In contrast to varying TiO$_2$ melt abundance and its effect on atmospheric TiO, increasing SiO$_2$ melt abundance does not lead to a direct increase of SiO or SiO$_2$ throughout the entire atmosphere. Lower in the atmosphere (at higher pressures) an increase in SiO$_2$ melt abundance does lead to an increase in atmospheric SiO and SiO$_2$. However, as we can see in panel \textit{e} (the TP profile of the planet orbiting a G-type star) in Figure \ref{fig:SiO2_results}, the changing melt composition affects the TP profile such that there is an increase in temperature in the upper-atmosphere ($<$1e-2 bar) with increasing SiO$_2$. This temperature increase is enough to cause SiO and SiO$_2$ to start dissociating into Si lower in the atmosphere. This causes the SiO and SiO$_2$ VMRs to be about equal for all melt abundances at pressures below 1e-3 and 1e-5 bar respectively. SiO and SiO$_2$ have the greatest amount of opacity contributions at greater pressures, so the spectra of these atmospheres still see significant changes. 

Increasing SiO$_2$ melt abundance leads to an increase in surface temperature for the planets around all four tested stellar spectral types. The temperature increase is caused by a relative decrease in TiO$_2$ in the melt as SiO$_2$ melt abundance is increased. SiO$_2$ makes up such a large fraction of the total melt composition ($\simeq45\%$ in BSE), that changing its abundance also significantly affects the abundances of the other melt species. The resulting decrease in TiO$_2$ abundance leads to an increase in temperature similar to that seen in the previous section where we decreased TiO$_2$ melt abundance directly.

Increasing the abundance of SiO$_2$ in the melt causes the temperature inversion to take place at higher atmospheric pressures. This is due to a combination of: 1) an increase in IR opacity due to more abundant atmospheric SiO$_2$ and SiO, and 2) a decrease in the optical opacity with less abundant atmospheric TiO. For the planets orbiting the F-, G-, and K-type, this moves the thermal inversion from around 1e-3 bar to around 1e-2 bar for a 10x increase in SiO$_2$. 

This change moves the temperature inversion from a region in the atmosphere above the SiO$_2$ opacity contribution to a region below it (see Figure \ref{fig:contribution_plot}), turning the SiO$_2$ absorption feature between 7 and 8 $\mu$m into an emission feature. This indicates that detecting an SiO$_2$ emission or absorption feature in the emission of a HRE could constrain the pressure/altitude at which a thermal inversion takes place.

Although the temperature inversion starts at a higher pressure and with a higher minimum temperature, the increase in IR absorption leads to the higher temperatures higher up in the atmosphere (as mentioned above as well). This causes the overall size of the temperature inversion to remain the same, but shifted towards higher temperatures for greater SiO$_2$ melt abundances. This, in combination with a greater abundance of SiO in the atmosphere, leads to very prominent SiO emission features (at 4.5 and 9 $mu$m) at high SiO$_2$ melt abundance which are up to 10 ppm greater for G-type stars and around 7 ppm greater for F and K type stars than that of the BSE composition spectra (Figure \ref{fig:SiO2_results}). 

For planets orbiting M-type stars (bottom row in Figure \ref{fig:SiO2_results}) we see different behavior than for the hotter spectral types. Just as what we saw for TiO$_2$ melt abundance variation, increasing SiO$_2$ melt abundance decreases the temperature inversion strength down to the point where it is almost non-existent and most of the atmosphere is non-inverted. This leads to the atmospheric SiO and SiO$_2$ features becoming absorption features instead of emission features. Although difficult to discern when overlaying the spectra over each other, it becomes clearer when we look at the flux difference between the BSE case and the increased SiO$_2$ melt abundance models. 

\begin{figure*}
    \centering
    \includegraphics[width=\linewidth]{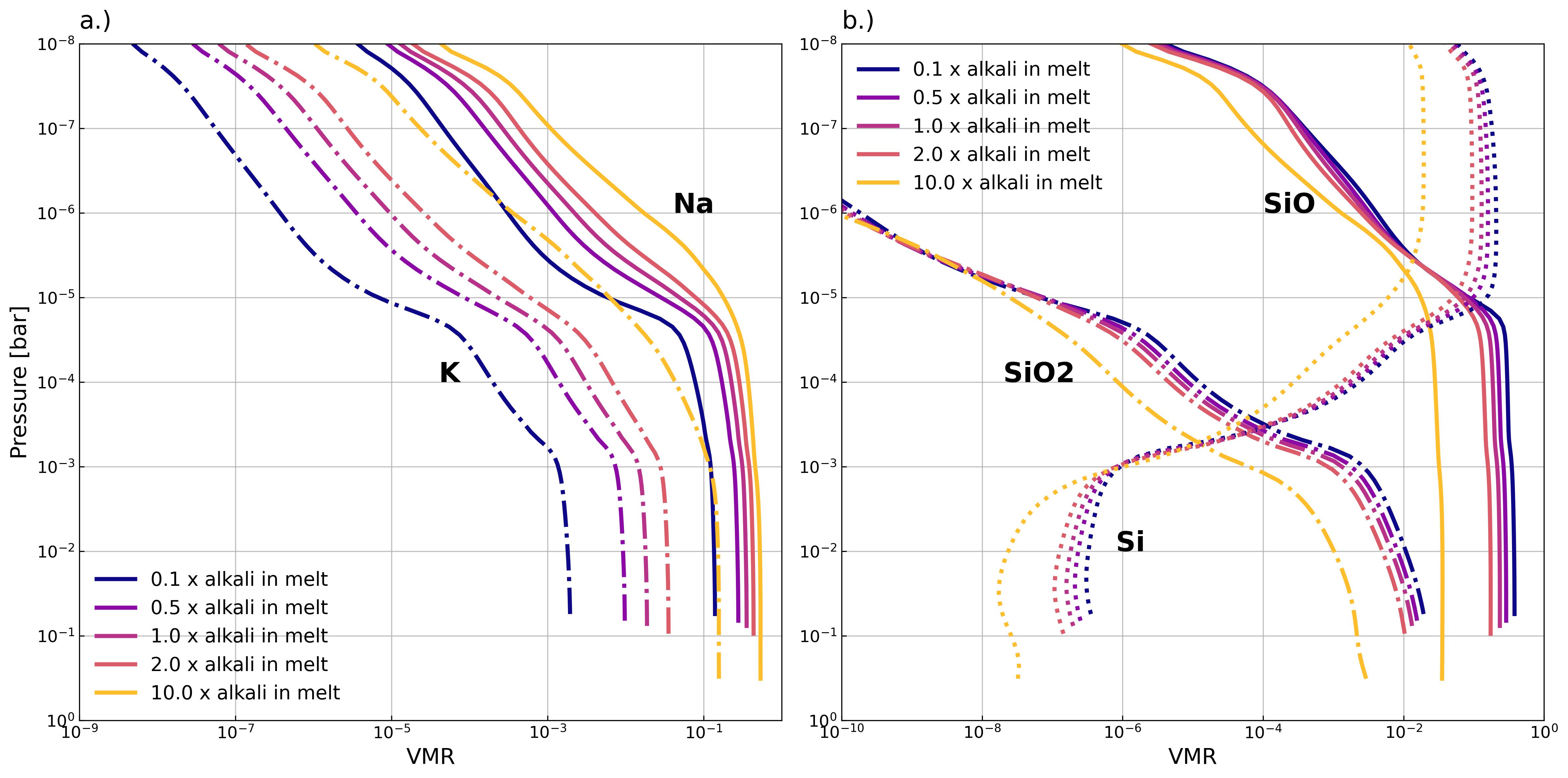}
    \caption{\textbf{Atmospheric species changing with alkali melt abundances:} Shown for a planet orbiting a G-type star (see Table \ref{tab:star_types} for parameters). The alkali oxide (Na$_2$O and K$_2$O) abundances in the melt were changed with respect to the BSE abundance (see Table \ref{tab:compositions}).}
    \label{fig:alkali_si_VMR}
\end{figure*}

\begin{figure*}
    \centering
    \includegraphics[width=0.85\linewidth]{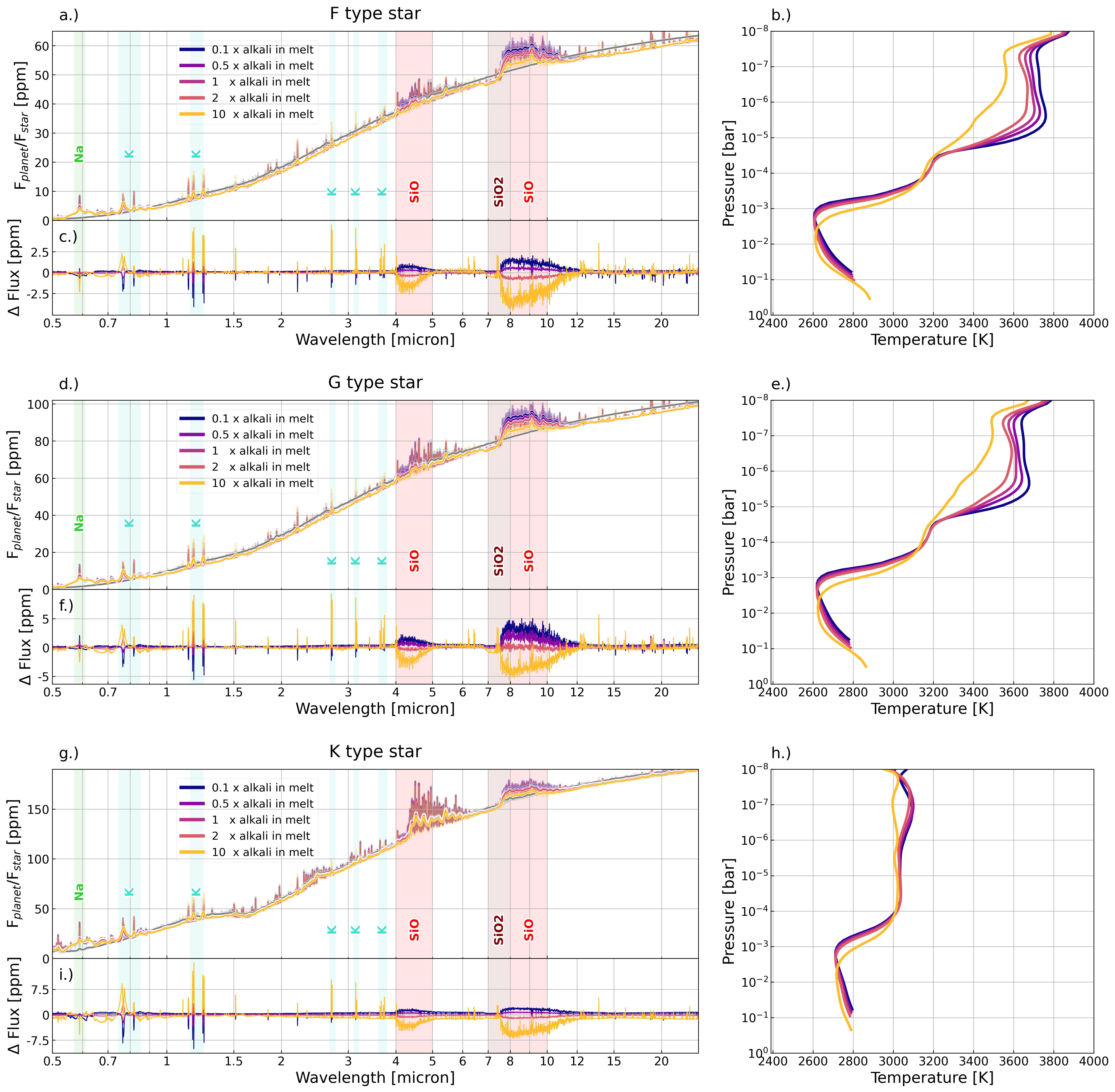}
    \caption{\textbf{Effect of varying Na$_2$O and K$_2$O melt percentage on emission spectra:} For an F-, G-, and K-type stars, we show the calculated pseudo-2D emission spectra and TP profiles for different SiO$_2$ melt abundances. Under each spectrum we also plot the differences in flux between the different spectra with respect to 1 x alkali abundances (which is the same as the Na$_2$O and K$_2$O weight percentages in a BSE composition). In grey we plotted the emission spectra of a model without any atmosphere for reference. The TP profiles shown are those for the models calculated at a 40$^\circ$ angle away from the substellar point. Note the difference in scales of the y-axes of the emission spectra for the different stellar types.}
    \label{fig:alkali_results}
\end{figure*}

\subsection{Alkali variation}

Atmospheric Na and K do not contribute as significantly over broad wavelengths to the opacity of the atmosphere as TiO, SiO, and SiO$_2$. Hence, they do not have as strong an influence on the TP profile or low-resolution spectrum of HRE exoplanet atmospheres. However, as can be seen in Figure \ref{fig:opacites}, they do have a number of lines across the optical and the near-infrared. They are also some of the most volatile species released in melt vaporisation from Na$_2$O and K$_2$O \citep{fegley_vaporization_1987, schaefer_thermodynamic_2004, zilinskas_observability_2022, van_buchem_lavatmos_2023, wolf_vaporock_2023}. As such, varying the abundance of Na$_2$O and K$_2$O in the melt has a strong effect on the atmospheric abundances of Na and K. We decided to vary Na$_2$O and K$_2$O abundance concurrently due to the similarity of their behavior. 

In Figure \ref{fig:alkali_si_VMR} we show the VMR of atmospheric Na and K (in panel \textit{a}) for different variations of the alkali oxide (Na$_2$O and K$_2$O) abundances in the melt at an angle of 40$^\circ$ away from the substellar point for a G-type star. Decreasing the alkali abundance by an order of magnitude relative to the BSE abundance (0.1x) leads to a drop in total surface pressure from 7.63e-2 bar to 5.52e-2 bar, while increasing the alkali abundance by an order of magnitude increases the total surface pressure to 3.21e-1 bar. Looking at Figure \ref{fig:alkali_results} - we see that this is not due to a change in surface temperature (as we saw for TiO$_2$ and SiO$_2$) but due to an increase in vaporised Na and K. Although Figure \ref{fig:alkali_si_VMR} focuses on a planet orbiting a G-type star, we see a similar increase in surface pressure for all host star spectral types, as can be seen in the TP profiles shown in Figure \ref{fig:alkali_results}. This could be of consequence to models that take heat and material redistribution over the day/night side of the planet into account \citep[e.g.][]{nguyen_clouds_2024}. 

Increasing the melt abundance of Na$_2$O and K$_2$O and the resulting increase in Na and K in the atmosphere leads to broadening of their features in the optical. The strongest effect is seen for the K features between 0.75 and 0.8 $\mu$m (see Figure \ref{fig:alkali_results}). There are also a number of emission lines between 1 and 3.5 $\mu$m from atmospheric K that see significant growth of (up to 10 ppm for planets orbiting a K-type star) with increasing alkali melt abundance - these are indicated in light blue in Figure \ref{fig:alkali_results}. Although the Na feature at 0.6 $\mu$m appears to broaden slightly, its peak decreases in height. This is likely due to our model not being of high enough spectral resolution to fully visualize this feature. Further work in high resolution is required to draw any definite conclusions.

Varying the abundance of alkali species in the melt significantly affects the SiO and SiO$_2$ atmospheric abundance, as shown in panel \textit{b} of Figure \ref{fig:alkali_si_VMR}. As a result, the SiO and SiO$_2$ features in the emission spectrum vary (see Figure \ref{fig:alkali_results}) and the temperature structure of the atmosphere is affected (most significantly above 1e-5 bar). This effect on the atmospheric Si species is partly due to the change in abundance of SiO$_2$ in the melt after renormalization (similar to the change in TiO$_2$ abundance when increasing SiO$_2$ abundance in the previous subsection). However, this change in abundance is relatively small due to the alkalis making up but a small percentage of the total melt composition. Besides melt abundance, the activity (see section \ref{subsec:equ_chemistry}) of a species in a melt is also determined by the activity coefficient ($\gamma$). In an ideal case, this is equal to one but due to the complex nature of melts this value is usually less than one. As shown in \cite{hirschmann_effect_1998}, increasing the alkali contents of a melt decreases the activity coefficient of SiO$_2$ due to reducing the number of Si-O-Si linkages. In Figure \ref{fig:sio2_activity} we plot both the activity of SiO$_2$ and its weight percentage in the melt, illustrating this effect. We do not see any significant changes in the spectrum of the M-type star and hence do not include it. 

\begin{figure}
    \centering
    \includegraphics[width=0.86\linewidth]{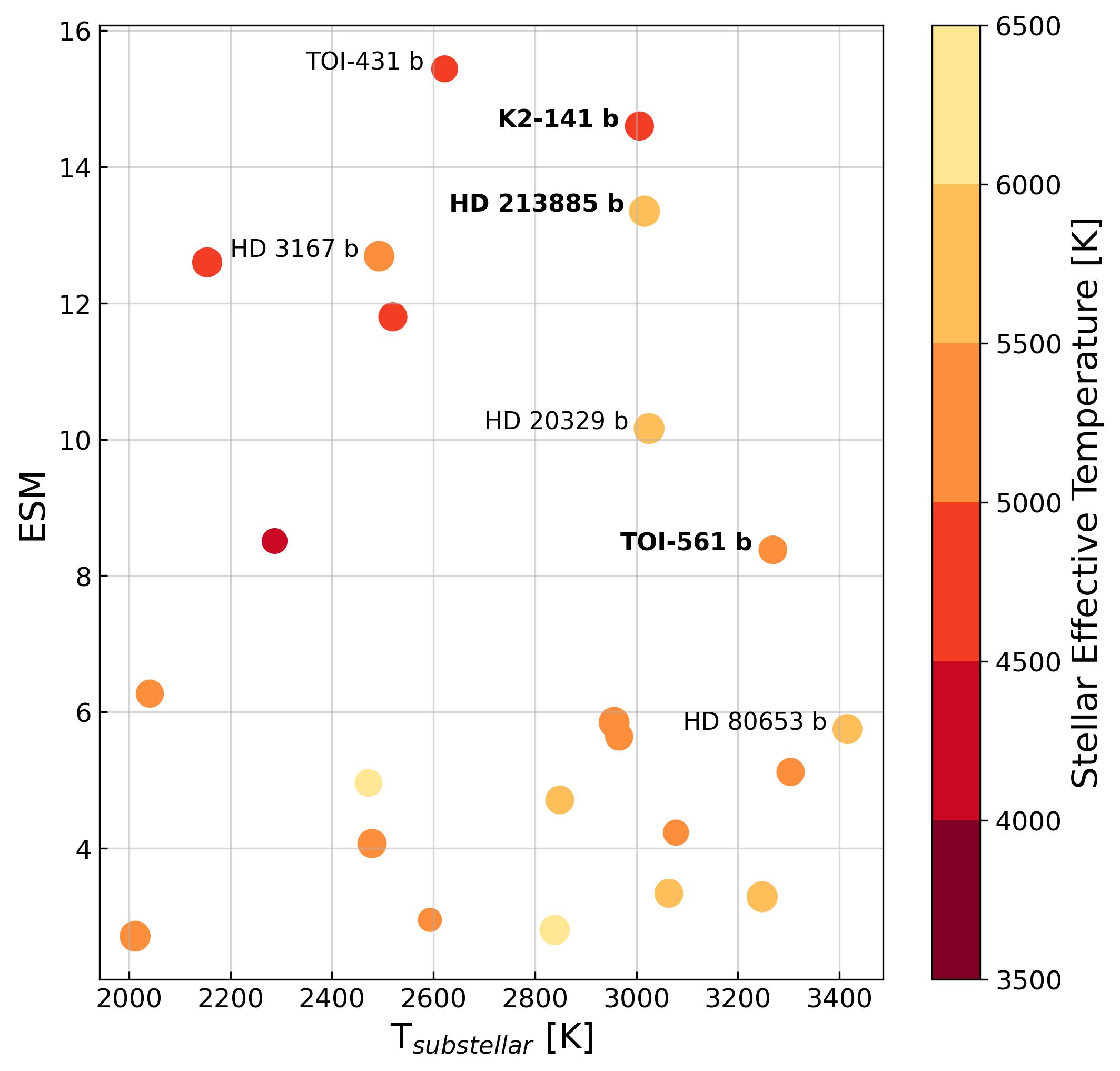}
    \caption{\textbf{Potential observing targets -} A selection of confirmed exoplanets with a radius of $<2R_{\oplus}$ and with a substellar equilibrium temperature $> 2000$ K. On the y-axis we plot the emission spectroscopy metric (ESM) as defined in \citet{kempton_framework_2018}. The planets labeled in bold are those we selected for error bar estimates shown in Figures \ref{fig:real_SiO2} and \ref{fig:real_TiO2}.} 
    \label{fig:esm_plot}
\end{figure}

\begin{figure*}
    \centering
    \includegraphics[width=0.86\linewidth]{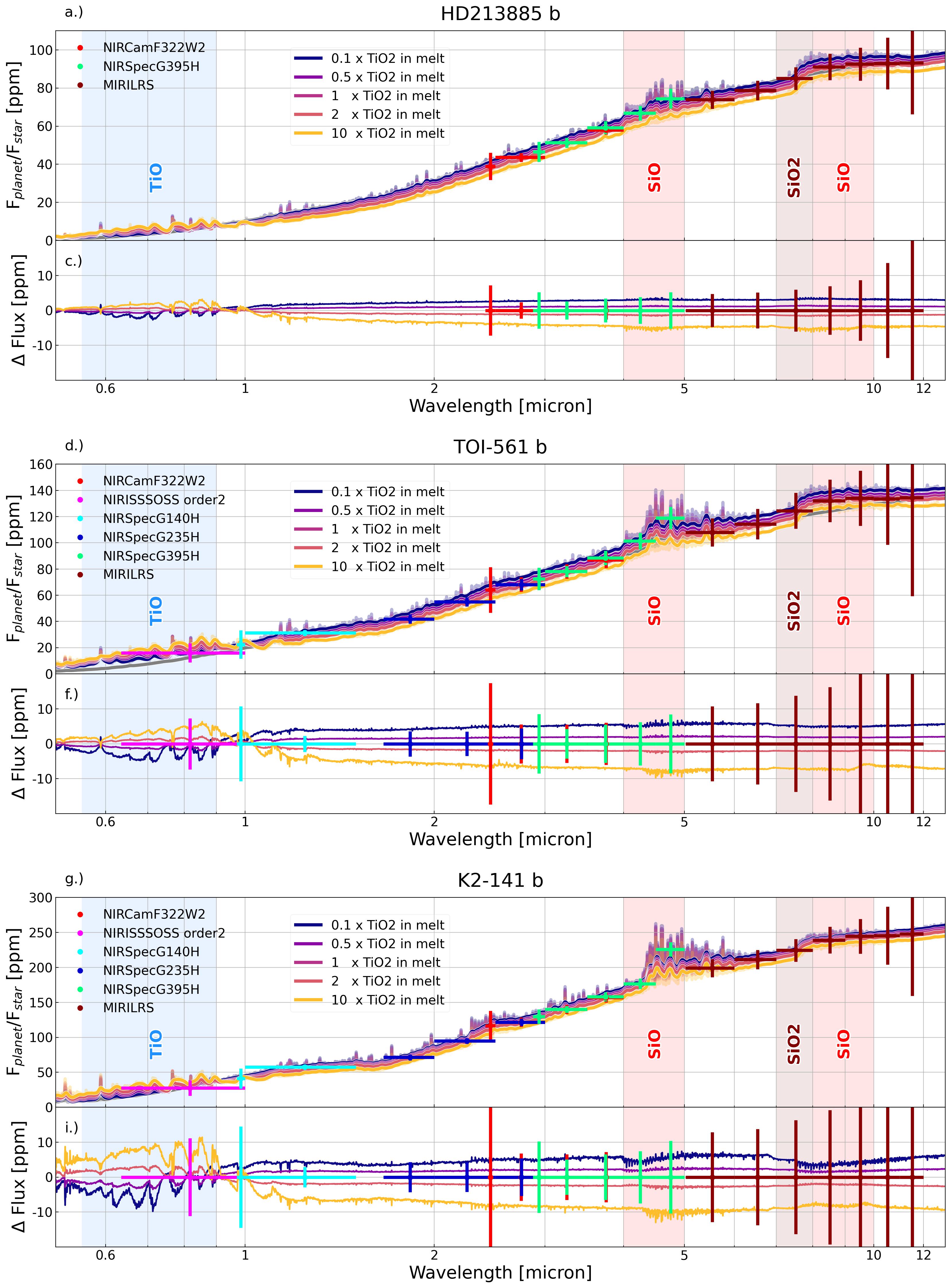}
    \caption{\textbf{JWST error bar estimates for TiO$_2$ variation -} Error estimates were calculated using \texttt{PANDEXO} \citep{batalha_pandexo_2017} for 12 eclipses.}
    \label{fig:real_TiO2}
\end{figure*}

\begin{figure*}
    \centering
    \includegraphics[width=0.86\linewidth]{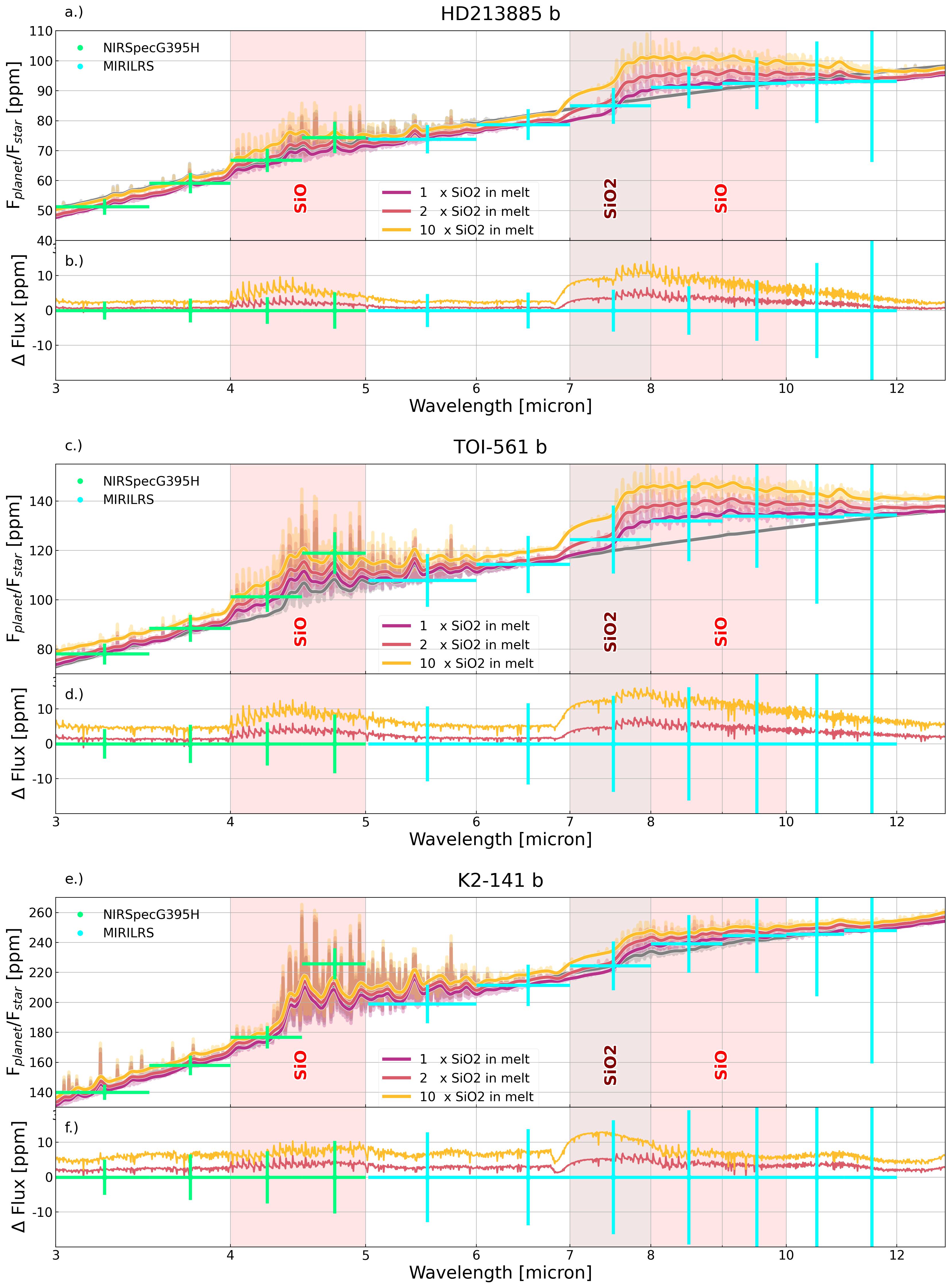}
    \caption{\textbf{JWST error bar estimates for SiO$_2$ variation -} Error estimates were calculated using \texttt{PANDEXO} \citep{batalha_pandexo_2017} for 12 eclipses.}
    \label{fig:real_SiO2}
\end{figure*}

\subsection{Observability of potential targets}

We simulated eclipse observations for three different HREs, with the aim of estimating the feasibility of differentiating different melt compositions from a BSE composition. To decide which HREs would be most suited for this, we made a selection of all confirmed HREs with a radius $<2R_{\oplus}$ and with a substellar equilibrium temperature $> 2000$ K. These are plotted in Figure \ref{fig:esm_plot} as a function of substellar temperature on the x-axis and their emission spectroscopy metric (ESM) on the y-axis. The ESM, as defined in \cite{kempton_framework_2018}, is a useful indicator of the expected S/N of a JWST secondary eclipse observation at mid-IR wavelengths. We colored the markers indicating the planets according to the effective temperature of their host stars. We selected HD 213885 b, TOI-561 b, and K2-141 b due to their high ESM values and substellar equilibrium temperatures, as well as the fact that their host stars cover a range of different stellar spectral types. For our simulated observations, we used the open source code \texttt{PANDEXO} \citep{batalha_pandexo_2017} and assumed 12 eclipses. We chose a number of groups per integration that is $\leq80\%$ of full well-depth in order to avoid reaching saturation. We assumed white noise and inversely scale the uncertainties for each of the instruments by the inverse square root of the number of observations.

In Figures \ref{fig:real_TiO2} and \ref{fig:real_SiO2}, we plot the synthetic emission spectra of the three selected HREs for varying TiO$_2$ and SiO$_2$ abundances as well as the estimated error bars for different JWST instruments based on the simulated observations of 12 eclipses. For HD 213885 b, we used the system parameter values published in \citet{espinoza_hd_2020}. Due to the brightness of the star HD 213885 (V$_{mag} = 7.95$, J$_{mag} = 6.81$), the number of instruments that can be used to observe it without over-saturating is limited, inhibiting observations below 2.5 $\mu$m. However, for the instruments with which we can observe the error bars are relatively small, $\sim2-3$ ppm at their best for NIRSpec G395H. For TOI-561 b and K2-141 b, we used the system parameters published in \citet{brinkman_toi-561_2022} and \citet{zieba_k2_2022} respectively.

To estimate the certainty with which we could potentially differentiate the different melt compositions from a BSE composition based on the corresponding emission spectra, we calculated the rejection significance of these spectra if one was to assume that the BSE composition is the ground truth. We do this by performing a $\chi^2$ test for which we calculated the survival function of the $\chi^2$ distribution and convert the resulting p-value into a Gaussian sigma rejection level by using the inverse cumulative distribution function. In Table \ref{tab:real_sigma} we show the certainties with which the spectra made using different melt compositions can be differentiated from the spectrum made using a BSE composition. For each planet, we use the simulated data of the instrument which provides the highest rejection significances.

\begin{table}
\caption{\textbf{Rejection significance of varying melt compositions} - Calculated with the assumption that BSE is the "ground-truth". These values allow us to estimate the confidence with which changes in the melt composition, with respect to BSE, can be discerned based on the emission spectrum as observed by the specified instruments after 12 eclipses.}
\label{tab:real_sigma}
\resizebox{\columnwidth}{!}{%
\begin{tabular}{cclccccc}
\multirow{2}{*}{\textbf{Planet}} & \multirow{2}{*}{\textbf{Instrument}} &  & \multicolumn{4}{c}{\textbf{M.C. Significance ($\sigma$)}} & \multirow{2}{*}{\textbf{\begin{tabular}[c]{@{}c@{}}Varied melt\\ oxide\end{tabular}}} \\
 &  &  & 0.1x & 0.5x & 2x & 10x &  \\ \hline
HD213885b & NIRCam F322W2 &  & 2.98 & 1.51 & 1.59 & 4.36 & \multirow{3}{*}{TiO$_2$} \\
K2-141b & NIRSpec G235H &  & 2.57 & 1.46 & 1.56 & 4.14 &  \\
TOI-561b & NIRSpec G235H &  & 3.14 & 1.52 & 1.57 & 3.99 &  \\ \hline
 &  &  &  &  &  &  &  \\
HD213885b & NIRCam F444W &  & - & - & 1.50 & 3.01 & \multirow{3}{*}{SiO$_2$} \\
K2-141b & NIRSpec G395H &  & - & - & 1.36 & 2.61 &  \\
TOI-561b & NIRSpec G395H &  & - & - & 1.33 & 2.96 & 
\end{tabular}%
}
\end{table}

In Table \ref{tab:real_sigma} the rejection significances of the varying TiO$_2$ and SiO$_2$ melt compositions are shown. For all three planets, 10x BSE TiO$_2$ melt abundance can be discerned with a relatively high certainty of 4 sigma from a regular BSE melt composition. A TiO$_2$ melt abundance of 0.1x can be discerned with a certainty of about 3 sigma for all three planets. 10x SiO$_2$ leads to less easily discernable changes, but could still be discerned from a BSE composition with a bit more than 3 sigma for HD213885b.

If these planets support dry-lava-vapor atmospheres and it is possible to observe them for 12 eclipses under favourable circumstances, these findings suggest that variations in SiO$_2$ and TiO$_2$ melt abundance of an order of magnitude away from that of BSE could potentially be discerned. However, similar changes in the flux as those due to TiO$_2$ variation, could also be caused by changes in the efficiency of heat-redistribution. Hence, the detection of the TiO feature at 0.5-1 $\mu$m may be necessary as extra evidence for a TiO$_2$ rich melt.

Another point to take into consideration is that the absence of a strong SiO or TiO feature may not be a full proof indication of a low SiO$_2$ or TiO$_2$ melt abundance respectively. If there were to exist a mechanism through which O$_2$ may be added or removed from the melt-atmosphere system in such a way that the free-parameter approach of modeling the melt vapor is more representative (see section \ref{subsec:o2_discussion}), then the absence of a strong SiO or TiO feature could also be an indication of a $\Delta\text{IW}>0$ \citep{seidler_impact_2024}.


\section{Discussion}\label{sec:discussion}

\subsection{Melt compositions}\label{sec:disc_melt_comp}

In section \ref{subsec:melt_comp}, we have shown (Figure \ref{fig:comp_distribution}) that the tested compositional ranges of individual silicate melt components mostly exceed the compositional variation of these components in bulk silicate reservoirs estimated from stellar compositions \citep{putirka_composition_2019}. However, studies of the early chemical evolution of rocky bodies in our own solar system show that the possible range in composition of lava oceans is likely to be significantly larger than the range in bulk silicate reservoir compositions. Here we focus specifically on variations in SiO$_2$ and TiO$_2$ abundances, as these show the highest sensitivity to atmospheric spectra.

The SiO$_2$ as well as TiO$_2$ content of magmas can vary significantly, both as a function of crust-mantle differentiation and crystal formation during magma cooling. To illustrate TiO$_2$ abundance variations, consider the chemical compositional evolution of magma and crust on the Moon. A wide range of geochemical evidence suggests that the Moon was fully molten shortly after its formation in the aftermath of a giant impact on Earth \citep[e.g.][]{steenstra_possible_2020}. At this time the lunar core segregated from the lunar mantle, and the composition of the molten surface was equal to the composition of the whole silicate reservoir, estimated to contain $\sim0.5$ wt.\% TiO$_2$. Cooling of the magma led to the crystallisation of a range of silicate minerals (olivine, pyroxenes) that do not contain TiO$_2$. As a result the magma composition evolved to contain $\sim1.6 \text{wt.}\%$ TiO$_2$ before any crust was formed on top of the magma \citep{lin_evidence_2017}. This abundance already exceeds the maximum TiO$_2$ content in star-derived bulk compositions by a factor of $\sim3$, see Figure \ref{fig:comp_distribution}.  Further cooling led to the formation of a > 30km thick primary crust formed primarily by the mineral plagioclase \citep[e.g.][]{elkins-tanton_lunar_2011,lin_evidence_2017}, which contains negligible TiO$_2$. The TiO$_2$ content of subsurface remaining magma exceeded 6 wt.\% at this point. Subsequent volcanism concentrated on the lunar nearside then led to the formation of a range of lavas with TiO$_2$ abundances up to $>15$ wt.\%, well beyond the maximum of the range of lava compositions based on bulk silicate abundance estimates (Figure \ref{fig:comp_distribution}).

A nice illustration of the large range of SiO$_2$ abundances in natural magma is provided by \cite{villiger_liquid_2004}. They experimentally studied the evolution of tholeiitic basalt, the most common basalt composition on Earth containing roughly 49 wt.\% SiO$_2$, equivalent to the maximum in the star-derived compositional range (\ref{fig:comp_distribution}), during magma cooling. Crystal fractionation during cooling leads to an increase in SiO$_2$ abundance up to 80 wt.\%, close to the maximum abundance considered in our models.

These examples illustrate that high TiO$_2$ and SiO$_2$ magma abundances, far exceeding star-derived estimates, occur in nature through regular geochemical processes - in this case through magma evolution and crystal segregation. In turn this implies that identifying such high abundances in magma would suggest significant magma evolution has taken place on an exoplanet, either through magma ocean cooling, or by remelting previously evolved solid reservoirs.

\subsection{Determining partial pressure of \texorpdfstring{O$_2$}{O2}}\label{subsec:o2_discussion}

LavAtmos \citep{van_buchem_lavatmos_2023}, the vaporisation model used in this work, assumes that the P$_{\text{O}_2}$ of the melt vapor can be determined by applying the "stoichiometric approach" of using the dual constraints of mass-action and mass-conservation, the same approach used by the \texttt{MAGMA} code \citep{fegley_vaporization_1987,schaefer_thermodynamic_2004,miguel_compositions_2011,kite_atmosphere-interior_2016,zilinskas_observability_2022}. As mentioned in section \ref{subsec:equ_chemistry}, this differs from the "free-parameter" approach used in the work of \cite{wolf_vaporock_2023} and \cite{seidler_impact_2024}. It remains a point of contention as to which approach is most suited for the modelling of melt-vapor. 

The free-parameter approach equates $P_{\text{O}_2}$ to the oxygen fugacity (fO$_2$) of the melt, which is then set according to the redox state of the melt, as approximated using (for example) the IW buffer - linking the fO$_2$ to the valence(s) of iron in the system. This allows one to assess the chemical compositions of a vapor above a melt for different oxidation states. \cite{seidler_impact_2024} make use of this to show how varying fO$_2$ over a wide range of values can change the composition of the vapor coming from the melt, the subsequent speciation of the atmosphere, and the resulting emission spectra. This is due to the strong dependence of most vaporisation reactions on the P$_{\text{O}_2}$ of the atmosphere overlying the melt. The stoichiometric approach on the other hand, is not very sensitive to changes in the iron valence state. This is because, depending on the total abundance of Fe species in the melt, the FeO and Fe$_2$O$_3$ vaporisation reactions contribute a marginal amount of O to the vapor, in contrast with more abundant or more volatile melt species such as SiO$_2$ and Na$_2$O. 

There are two important drawbacks to consider when not constraining $P_{\text{O}_2}$ using mass action and mass balance. First, imposing a $P_{\text{O}_2}$ value other than that which satisfies mass action means that oxygen is being added or removed from the system - depending on if this partial-pressure is higher or lower than the value that satisfies these constraints. This open system approach essentially treats oxygen as a perfectly mobile component \citep[e.g.][]{korzhinskii_physicochemical_1959}, and assumes that are always sufficient sources and sinks for O available in the immediate surrounds of the magma \citep[e.g.][]{ghiorso_chemical_1985-1}. This may be a fine assumption for smaller magma bodies such as magmatic intrusions on Earth, but whether it is valid for the magma ocean settings considered here is unclear. Second, it does not allow for $P_{\text{O}_2}$ to change self-consistently as the composition of the melt is changed. With LavAtmos, we can consistently account for these changes when testing different compositions.

\citet{seidler_impact_2024} argue that using the stoichiometric approach to calculate $P_{\text{O}_2}$ is only valid if all possible vaporisation reactions that could release O$_2$ are taken into account and if the set of melt components and their thermodynamic properties is complete and accurate. For the nine included end-member components of the melt, LavAtmos includes 30 of the most important vaporisation reactions, of which 26 involve O$_2$ \citep{van_buchem_lavatmos_2023}. At 4000 K for a BSE melt composition, the partial pressure of the least abundant vapor species (Al$_2$) is just above 1e-9 bar, while the partial pressure of O$_2$ is just under 1 bar. At colder temperatures, the difference between these partial pressures grows further. Including the vaporisation reactions of the less abundant vapor species would likely not significantly alter the calculated O$_2$ partial pressure. Similarly, including a greater number of refractory oxide species in the thermodynamic melt model would likely only have a marginal impact on the calculated O$_2$ partial pressure. Concerning the accuracy of the thermodynamic properties, we agree that this is something that should be looked into - however, since both Vaporock and \texttt{MAGMA} also make extensive use of the JANAF tables this issue is not unique to LavAtmos. Hence, in our view, LavAtmos provides an approximation of the vapor composition that is sufficiently accurate for the work presented in this paper.

The free parameter approach provides a direct way to alter the amount of oxygen in the system, with the drawback of not adhering to mass conservation. The stoichiometric approach adheres to mass conservation, but does not provide a straightforward manner of changing the amount of O in the system. This, in part, disconnects the vapor composition from the redox state of the melt. This distinction between the two approaches is important for interpreting and/or reproducing our findings. Until a new vaporisation model is developed that is able to reconcile these two approaches, it is our view that the appropriateness of the approaches should be judged on a case by case basis depending on the aims and needs of the modelling work being done.

\subsection{Further assumptions made in our model}

The model used for this work is a pseudo-2D model that assumes thermochemical equilibrium (see section \ref{sec:methods}). This does not take into account chemical kinetics that may be taking place higher up in the atmosphere where strong UV radiation could lead to chemical dissociation. Due to the high temperatures of the modeled atmospheres, it is likely that a chemical equilibrium approach provides a good approximation of the atmospheric chemistry. In order to investigate whether photochemistry could have a significant influence, chemical pathways for silicates are needed, which are currently unavailable in literature. 

We do not take condensation into account either. As shown in \citet{nguyen_clouds_2024}, this could have some effect on the temperature structure of the planet at different longitudes. Future work could be focused on combining condensation calculations with a pseudo-2D model to take the resulting changes in temperature structure and (potentially) cloud opacity into account. 

The fact that we focus on low resolution spectra means that our results do not rule out that other melt oxide species could be important to high resolution spectral analysis. 

\subsection{Surface temperature degeneracy} 

As discussed in the results section and as shown in Figures \ref{fig:TiO2_results} and \ref{fig:real_TiO2}, varying TiO$_2$ melt abundance, directly influences TiO atmospheric abundance, which has a very strong influence on the surface temperature of the planet. This is due to its strong optical opacity. 

Concluding that a HRE atmosphere contains TiO because the apparent surface temperature is lower than what one would expect without TiO, would be ignoring other key factors that can affect planetary surface temperatures and should therefore not be used as direct evidence. For example, in \citet{hu_secondary_2024}, the emission spectrum found for 55-Cnc e pointed to a surface temperature of several hundreds of Kelvin lower than expected for a bare lava-vapor atmosphere. This was attributed to the presence of a volatile (potentially carbon-rich) atmosphere leading to more efficient heat-redistribution from the day- to the night-side. Other studies have shown that the presence of clouds may have a significant effect on surface temperatures as well \citep{nguyen_clouds_2024,loftus_extreme_2024}. 

Conversely, future interpretations of HRE emission spectra should also consider the effect that a significant TiO abundance in the atmosphere could have on the temperature-pressure structure of a lava planet. A potential way to disentangle the resulting degeneracy would be to try and observe the TiO emission feature found between 0.55 and 0.9 $\mu$m.
    
\section{Conclusions}\label{sec:conclusion}

We investigated how varying the abundances of oxide melt species at the surface of a lava planet affects its atmospheric emission spectrum. We modelled the atmospheres and resulting emission spectra using chemically self-consistent pseudo-2D models for dry HREs orbiting F-, G-, K-, and M-type host stars. We find that the melt species which have the most significant influence on the emission spectra of HREs are TiO$_2$ and SiO$_2$.

TiO$_2$ melt abundance dictates atmospheric TiO abundance, which, due to its strong optical opacity, heavily influences the surface temperature of the planet if it orbits an F-, G-, or K- type star. This change in surface temperature affects both the total pressure of the vaporised atmosphere and its composition - affecting the strength of emission features of species such as SiO. The decrease in temperature caused by the presence of TiO affects all but the upper reaches of the atmosphere, leading to a drop in the continuum of emission of the planet. A potential way to discern this drop in temperature as being due to the presence of TiO and not due to efficient heat-redistribution is by observing the TiO optical emission feature. 

Increasing the SiO$_2$ melt abundance leads to a significant increase in the strength of the atmospheric SiO and SiO$_2$ features in the infrared for planets orbiting F-, G-, and K-type stars. It also leads to stronger and deeper temperature inversions, below the point where SiO$_2$ has the highest opacity contribution. This leads to the SiO$_2$ feature flipping from being an absorption feature to being an emission feature.

Varying the melt abundance of the alkali oxides Na$_2$O and K$_2$O influences the abundance of atmospheric Na and K respectively. This affects the K (and single Na) spectral features in the optical and near-infrared, which may merit further investigation with high-resolution modelling for HRE planets orbiting F-, G-, or M-type stars. The increase in atmospheric Na and K at relatively large melt abundances of Na$_2$O and K$_2$O lead to a large increase in the total pressure of the atmosphere. This could potentially be of importance for heat-redistribution models. 

Varying the alkali abundance in the melt also affects the activity of the SiO$_2$ in the melt, with greater alkali abundances leading to lower SiO$_2$ activities. This significantly impacts the abundances of atmospheric Si species and hence the strength of the SiO and SiO$_2$ features in the emission spectrum.  

To estimate the observability of these changes in emission spectra due to changes in melt composition, simulated JWST observations for HD213885 b, TOI-561 b, and K2-141 b. These planets were selected due to their high infrared observability metric v  vand high surface temperatures. We found that with 12 eclipses, 0.1x and 10x variations in TiO$_2$ abundance in the melt may be discernable from a BSE composition with a certainty of about 3 and 4 sigma respectively for all three tested planets. 10x SiO$2$ abundance may be discernable with a certainty between 2.6 and 3 sigma.

\section*{Acknowledgements}

This work was supported financially through a Dutch Research Council (NWO) Planetary and Exoplanetary Science (PEPSci) grant awarded to Y.M. and W. van W. 
Y.M and M.Z. acknowledge funding from the European Research Council (ERC) under the European Union’s Horizon 2020 research and innovation programme (grant agreement no. 101088557, N-GINE). R. B. acknowledges funding from the Oort and Legaat Kruytbosch Scholarship. 

\section*{Data Availability}

The data underlying this article and that support the findings of this study are available upon reasonable request to the corresponding author.



\bibliographystyle{mnras}
\bibliography{references} 

@book{korzhinskii_physicochemical_1959,
	title = {Physicochemical basis of the analysis of the paragenesis of minerals},
	url = {http://archive.org/details/physicochemicalb0000korz},
	language = {eng},
	urldate = {2025-10-26},
	publisher = {New York, NY : Consultants Bureau},
	author = {Korzhinskii, Dmitri Sergeevich},
	collaborator = {{Internet Archive}},
	year = {1959},
}

@article{villiger_liquid_2004,
	title = {The {Liquid} {Line} of {Descent} of {Anhydrous}, {Mantle}-{Derived}, {Tholeiitic} {Liquids} by {Fractional} and {Equilibrium} {Crystallization}—an {Experimental} {Study} at 1·0 {GPa}},
	volume = {45},
	issn = {0022-3530},
	url = {https://doi.org/10.1093/petrology/egh042},
	doi = {10.1093/petrology/egh042},
	number = {12},
	urldate = {2025-10-26},
	journal = {Journal of Petrology},
	author = {Villiger, Samuel and Ulmer, Peter and Müntener, Othmar and Thompson, Alan Bruce},
	month = dec,
	year = {2004},
	pages = {2369--2388},
}

@article{elkins-tanton_magma_2012,
	title = {Magma {Oceans} in the {Inner} {Solar} {System}},
	volume = {40},
	issn = {0084-6597, 1545-4495},
	url = {https://www.annualreviews.org/content/journals/10.1146/annurev-earth-042711-105503},
	doi = {10.1146/annurev-earth-042711-105503},
	language = {en},
	number = {Volume 40, 2012},
	urldate = {2025-10-26},
	journal = {Annual Review of Earth and Planetary Sciences},
	author = {Elkins-Tanton, Linda T.},
	month = may,
	year = {2012},
	note = {Publisher: Annual Reviews},
	pages = {113--139},
}

@article{steenstra_geochemical_2020,
	title = {Geochemical constraints on core-mantle differentiation in {Mercury} and the aubrite parent body},
	volume = {340},
	issn = {0019-1035},
	url = {https://www.sciencedirect.com/science/article/pii/S0019103519303902},
	doi = {10.1016/j.icarus.2020.113621},
	urldate = {2025-10-26},
	journal = {Icarus},
	author = {Steenstra, E. S. and van Westrenen, W.},
	month = apr,
	year = {2020},
	pages = {113621},
}

@article{rai_core-mantle_2013,
	title = {Core-mantle differentiation in {Mars}},
	volume = {118},
	copyright = {©2013. American Geophysical Union. All Rights Reserved.},
	issn = {2169-9100},
	url = {https://onlinelibrary.wiley.com/doi/abs/10.1002/jgre.20093},
	doi = {10.1002/jgre.20093},
	language = {en},
	number = {6},
	urldate = {2025-10-26},
	journal = {Journal of Geophysical Research: Planets},
	author = {Rai, Nachiketa and van Westrenen, Wim},
	year = {2013},
	note = {\_eprint: https://agupubs.onlinelibrary.wiley.com/doi/pdf/10.1002/jgre.20093},
	pages = {1195--1203},
}

@article{badro_core_2015,
	title = {Core formation and core composition from coupled geochemical and geophysical constraints},
	volume = {112},
	url = {https://www.pnas.org/doi/10.1073/pnas.1505672112},
	doi = {10.1073/pnas.1505672112},
	number = {40},
	urldate = {2025-10-26},
	journal = {Proceedings of the National Academy of Sciences},
	author = {Badro, James and Brodholt, John P. and Piet, Hélène and Siebert, Julien and Ryerson, Frederick J.},
	month = oct,
	year = {2015},
	note = {Publisher: Proceedings of the National Academy of Sciences},
	pages = {12310--12314},
}

@article{wade_core_2005,
	title = {Core formation and the oxidation state of the {Earth}},
	volume = {236},
	issn = {0012-821X},
	url = {https://www.sciencedirect.com/science/article/pii/S0012821X05003286},
	doi = {10.1016/j.epsl.2005.05.017},
	number = {1},
	urldate = {2025-10-26},
	journal = {Earth and Planetary Science Letters},
	author = {Wade, J. and Wood, B. J.},
	month = jul,
	year = {2005},
	pages = {78--95},
}

@article{van_buchem_lavatmos_2025,
	title = {{LavAtmos} 2.0 - {Incorporating} volatile species in vaporisation models},
	volume = {695},
	copyright = {© The Authors 2025},
	issn = {0004-6361, 1432-0746},
	url = {https://www.aanda.org/articles/aa/abs/2025/03/aa50992-24/aa50992-24.html},
	doi = {10.1051/0004-6361/202450992},
	language = {en},
	urldate = {2025-10-05},
	journal = {Astronomy \& Astrophysics},
	author = {van Buchem, C. P. A. and Zilinskas, M. and Miguel, Y. and Westrenen, W. van},
	month = mar,
	year = {2025},
	note = {Publisher: EDP Sciences},
	pages = {A154},
}

@article{zilinskas_characterising_2025,
	title = {Characterising the atmosphere of 55 {Cancri} e: {1D} forward model grid for current and future {JWST} observations},
	volume = {697},
	copyright = {https://creativecommons.org/licenses/by/4.0},
	issn = {0004-6361, 1432-0746},
	shorttitle = {Characterising the atmosphere of 55 {Cancri} e},
	url = {https://www.aanda.org/10.1051/0004-6361/202554062},
	doi = {10.1051/0004-6361/202554062},
	language = {en},
	urldate = {2025-10-04},
	journal = {Astronomy \& Astrophysics},
	author = {Zilinskas, M. and Van Buchem, C. P. A. and Zieba, S. and Miguel, Y. and Sandford, E. and Hu, R. and Patel, J. A. and Bello-Arufe, A. and Janssen, L. J. and Tsai, S.-M. and Dragomir, D. and Zhang, M.},
	month = may,
	year = {2025},
	pages = {A34},
}

@article{patel_jwst_2024,
	title = {{JWST} reveals the rapid and strong day-side variability of 55 {Cancri} e},
	volume = {690},
	copyright = {© The Authors 2024},
	issn = {0004-6361, 1432-0746},
	url = {https://www.aanda.org/articles/aa/abs/2024/10/aa50748-24/aa50748-24.html},
	doi = {10.1051/0004-6361/202450748},
	language = {en},
	urldate = {2025-10-04},
	journal = {Astronomy \& Astrophysics},
	author = {Patel, J. A. and Brandeker, A. and Kitzmann, D. and Roche, D. J. M. Petit dit de la and Bello-Arufe, A. and Heng, K. and Valdés, E. Meier and Persson, C. M. and Zhang, M. and Demory, B.-O. and Bourrier, V. and Deline, A. and Ehrenreich, D. and Fridlund, M. and Hu, R. and Lendl, M. and Oza, A. V. and Alibert, Y. and Hooton, M. J.},
	month = oct,
	year = {2024},
	note = {Publisher: EDP Sciences},
	pages = {A159},
}

@article{seidler_impact_2024,
	title = {Impact of oxygen fugacity on the atmospheric structure and emission spectra of ultra-hot rocky exoplanets},
	volume = {691},
	copyright = {https://creativecommons.org/licenses/by/4.0},
	issn = {0004-6361, 1432-0746},
	url = {https://www.aanda.org/10.1051/0004-6361/202450546},
	doi = {10.1051/0004-6361/202450546},
	language = {en},
	urldate = {2025-10-04},
	journal = {Astronomy \& Astrophysics},
	author = {Seidler, Fabian L. and Sossi, Paolo A. and Grimm, Simon L.},
	month = nov,
	year = {2024},
	pages = {A159},
}

@article{zilinskas_observability_2023,
	title = {Observability of silicates in volatile atmospheres of super-{Earths} and sub-{Neptunes} - {Exploring} the edge of the evaporation desert},
	volume = {671},
	copyright = {© The Authors 2023},
	issn = {0004-6361, 1432-0746},
	url = {https://www.aanda.org/articles/aa/abs/2023/03/aa45521-22/aa45521-22.html},
	doi = {10.1051/0004-6361/202245521},
	language = {en},
	urldate = {2025-02-15},
	journal = {Astronomy \& Astrophysics},
	author = {Zilinskas, M. and Miguel, Y. and Buchem, C. P. A. van and Snellen, I. a. G.},
	month = mar,
	year = {2023},
	note = {Publisher: EDP Sciences},
	pages = {A138},
}

@article{zieba_k2_2022,
	title = {K2 and {Spitzer} phase curves of the rocky ultra-short-period planet {K2}-141 b hint at a tenuous rock vapor atmosphere},
	volume = {664},
	copyright = {© S. Zieba et al. 2022},
	issn = {0004-6361, 1432-0746},
	url = {https://www.aanda.org/articles/aa/abs/2022/08/aa42912-21/aa42912-21.html},
	doi = {10.1051/0004-6361/202142912},
	language = {en},
	urldate = {2025-02-15},
	journal = {Astronomy \& Astrophysics},
	author = {Zieba, S. and Zilinskas, M. and Kreidberg, L. and Nguyen, T. G. and Miguel, Y. and Cowan, N. B. and Pierrehumbert, R. and Carone, L. and Dang, L. and Hammond, M. and Louden, T. and Lupu, R. and Malavolta, L. and Stevenson, K. B.},
	month = aug,
	year = {2022},
	note = {Publisher: EDP Sciences},
	pages = {A79},
}

@article{kitzmann_fastchem_2024,
	title = {{FastChem} {Cond}: equilibrium chemistry with condensation and rainout for cool planetary and stellar environments},
	volume = {527},
	issn = {0035-8711},
	shorttitle = {fastchem cond},
	url = {https://doi.org/10.1093/mnras/stad3515},
	doi = {10.1093/mnras/stad3515},
	number = {3},
	urldate = {2025-02-15},
	journal = {Monthly Notices of the Royal Astronomical Society},
	author = {Kitzmann, Daniel and Stock, Joachim W and Patzer, A Beate C},
	month = jan,
	year = {2024},
	pages = {7263--7283},
}

@article{hirschmann_effect_1998,
	title = {The {Effect} of {Alkalis} on the {Silica} {Content} of {Mantle}-{Derived} {Melts}},
	volume = {62},
	copyright = {https://www.elsevier.com/tdm/userlicense/1.0/},
	issn = {00167037},
	url = {https://linkinghub.elsevier.com/retrieve/pii/S0016703798000283},
	doi = {10.1016/S0016-7037(98)00028-3},
	language = {en},
	number = {5},
	urldate = {2025-01-30},
	journal = {Geochimica et Cosmochimica Acta},
	author = {Hirschmann, M.M. and Baker, M.B. and Stolper, E.M.},
	month = mar,
	year = {1998},
	pages = {883--902},
}

@article{kempton_framework_2018,
	title = {A {Framework} for {Prioritizing} the {TESS} {Planetary} {Candidates} {Most} {Amenable} to {Atmospheric} {Characterization}},
	volume = {130},
	issn = {0004-6280, 1538-3873},
	url = {http://arxiv.org/abs/1805.03671},
	doi = {10.1088/1538-3873/aadf6f},
	language = {en},
	number = {993},
	urldate = {2024-11-21},
	journal = {Publications of the Astronomical Society of the Pacific},
	author = {Kempton, Eliza M.-R. and Bean, Jacob L. and Louie, Dana R. and Deming, Drake and Koll, Daniel D. B. and Mansfield, Megan and Christiansen, Jessie L. and Lopez-Morales, Mercedes and Swain, Mark R. and Zellem, Robert T. and Ballard, Sarah and Barclay, Thomas and Barstow, Joanna K. and Batalha, Natasha E. and Beatty, Thomas G. and Berta-Thompson, Zach and Birkby, Jayne and Buchhave, Lars A. and Charbonneau, David and Cowan, Nicolas B. and Crossfield, Ian and Val-Borro, Miguel de and Doyon, Rene and Dragomir, Diana and Gaidos, Eric and Heng, Kevin and Hu, Renyu and Kane, Stephen R. and Kreidberg, Laura and Mallonn, Matthias and Morley, Caroline V. and Narita, Norio and Nascimbeni, Valerio and Palle, Enric and Quintana, Elisa V. and Rauscher, Emily and Seager, Sara and Shkolnik, Evgenya L. and Sing, David K. and Sozzetti, Alessandro and Stassun, Keivan G. and Valenti, Jeff A. and Essen, Carolina von},
	month = nov,
	year = {2018},
	note = {arXiv:1805.03671 [astro-ph]},
	pages = {114401},
}

@misc{loftus_extreme_2024,
	title = {Extreme {Weather} {Variability} on {Hot} {Rocky} {Exoplanet} 55 {Cancri} e {Explained} by {Magma} {Temperature}-{Cloud} {Feedback}},
	url = {http://arxiv.org/abs/2409.16270},
	language = {en},
	urldate = {2024-11-13},
	publisher = {arXiv},
	author = {Loftus, Kaitlyn and Luo, Yangcheng and Fan, Bowen and Kite, Edwin S.},
	month = sep,
	year = {2024},
	note = {arXiv:2409.16270 [astro-ph]},
}

@misc{nguyen_clouds_2024,
	title = {Clouds on partial atmospheres of lava planets and where to find them},
	url = {http://arxiv.org/abs/2407.21111},
	doi = {10.48550/arXiv.2407.21111},
	urldate = {2024-11-07},
	publisher = {arXiv},
	author = {Nguyen, T. Giang and Cowan, Nicolas B. and Dang, Lisa},
	month = jul,
	year = {2024},
	note = {arXiv:2407.21111},
}

@article{espinoza_hd_2020,
	title = {{HD} 213885b: a transiting 1-d-period super-{Earth} with an {Earth}-like composition around a bright ({V} = 7.9) star unveiled by {TESS}},
	volume = {491},
	issn = {0035-8711},
	shorttitle = {{HD} 213885b},
	url = {https://ui.adsabs.harvard.edu/abs/2020MNRAS.491.2982E},
	doi = {10.1093/mnras/stz3150},
	urldate = {2024-11-07},
	journal = {Monthly Notices of the Royal Astronomical Society},
	author = {Espinoza, Néstor and Brahm, Rafael and Henning, Thomas and Jordán, Andrés and Dorn, Caroline and Rojas, Felipe and Sarkis, Paula and Kossakowski, Diana and Schlecker, Martin and Díaz, Matías R. and Jenkins, James S. and Aguilera-Gomez, Claudia and Jenkins, Jon M. and Twicken, Joseph D. and Collins, Karen A. and Lissauer, Jack and Armstrong, David J. and Adibekyan, Vardan and Barrado, David and Barros, Susana C. C. and Battley, Matthew and Bayliss, Daniel and Bouchy, François and Bryant, Edward M. and Cooke, Benjamin F. and Demangeon, Olivier D. S. and Dumusque, Xavier and Figueira, Pedro and Giles, Helen and Lillo-Box, Jorge and Lovis, Christophe and Nielsen, Louise D. and Pepe, Francesco and Pollacco, Don and Santos, Nuno C. and Sousa, Sergio G. and Udry, Stéphane and Wheatley, Peter J. and Turner, Oliver and Marmier, Maxime and Ségransan, Damien and Ricker, George and Latham, David and Seager, Sara and Winn, Joshua N. and Kielkopf, John F. and Hart, Rhodes and Wingham, Geof and Jensen, Eric L. N. and Hełminiak, Krzysztof G. and Tokovinin, A. and Briceño, C. and Ziegler, Carl and Law, Nicholas M. and Mann, Andrew W. and Daylan, Tansu and Doty, John P. and Guerrero, Natalia and Boyd, Patricia and Crossfield, Ian and Morris, Robert L. and Henze, Christopher E. and Chacon, Aaron Dean},
	month = jan,
	year = {2020},
	note = {Publisher: OUP
ADS Bibcode: 2020MNRAS.491.2982E},
	pages = {2982--2999},
}

@article{lothringer_influence_2019,
	title = {The {Influence} of {Host} {Star} {Spectral} {Type} on {Ultra}-hot {Jupiter} {Atmospheres}},
	volume = {876},
	issn = {0004-637X},
	url = {https://dx.doi.org/10.3847/1538-4357/ab1485},
	doi = {10.3847/1538-4357/ab1485},
	language = {en},
	number = {1},
	urldate = {2024-11-05},
	journal = {The Astrophysical Journal},
	author = {Lothringer, Joshua D. and Barman, Travis},
	month = may,
	year = {2019},
	note = {Publisher: The American Astronomical Society},
	pages = {69},
}

@article{loyd_muscles_2016,
	title = {{THE} {MUSCLES} {TREASURY} {SURVEY}. {III}. {X}-{RAY} {TO} {INFRARED} {SPECTRA} {OF} 11 {M} {AND} {K} {STARS} {HOSTING} {PLANETS}},
	volume = {824},
	issn = {0004-637X},
	url = {https://dx.doi.org/10.3847/0004-637X/824/2/102},
	doi = {10.3847/0004-637X/824/2/102},
	language = {en},
	number = {2},
	urldate = {2024-10-18},
	journal = {The Astrophysical Journal},
	author = {Loyd, R. O. P. and France, Kevin and Youngblood, Allison and Schneider, Christian and Brown, Alexander and Hu, Renyu and Linsky, Jeffrey and Froning, Cynthia S. and Redfield, Seth and Rugheimer, Sarah and Tian, Feng},
	month = jun,
	year = {2016},
	note = {Publisher: The American Astronomical Society},
	pages = {102},
}

@article{youngblood_muscles_2016,
	title = {{THE} {MUSCLES} {TREASURY} {SURVEY}. {II}. {INTRINSIC} {LYα} {AND} {EXTREME} {ULTRAVIOLET} {SPECTRA} {OF} {K} {AND} {M} {DWARFS} {WITH} {EXOPLANETS}*},
	volume = {824},
	issn = {0004-637X},
	url = {https://dx.doi.org/10.3847/0004-637X/824/2/101},
	doi = {10.3847/0004-637X/824/2/101},
	language = {en},
	number = {2},
	urldate = {2024-10-18},
	journal = {The Astrophysical Journal},
	author = {Youngblood, Allison and France, Kevin and Loyd, R. O. Parke and Linsky, Jeffrey L. and Redfield, Seth and Schneider, P. Christian and Wood, Brian E. and Brown, Alexander and Froning, Cynthia and Miguel, Yamila and Rugheimer, Sarah and Walkowicz, Lucianne},
	month = jun,
	year = {2016},
	note = {Publisher: The American Astronomical Society},
	pages = {101},
}

@article{france_muscles_2016,
	title = {{THE} {MUSCLES} {TREASURY} {SURVEY}. {I}. {MOTIVATION} {AND} {OVERVIEW}*},
	volume = {820},
	issn = {0004-637X},
	url = {https://dx.doi.org/10.3847/0004-637X/820/2/89},
	doi = {10.3847/0004-637X/820/2/89},
	language = {en},
	number = {2},
	urldate = {2024-10-18},
	journal = {The Astrophysical Journal},
	author = {France, Kevin and Loyd, R. O. Parke and Youngblood, Allison and Brown, Alexander and Schneider, P. Christian and Hawley, Suzanne L. and Froning, Cynthia S. and Linsky, Jeffrey L. and Roberge, Aki and Buccino, Andrea P. and Davenport, James R. A. and Fontenla, Juan M. and Kaltenegger, Lisa and Kowalski, Adam F. and Mauas, Pablo J. D. and Miguel, Yamila and Redfield, Seth and Rugheimer, Sarah and Tian, Feng and Vieytes, Mariela C. and Walkowicz, Lucianne M. and Weisenburger, Kolby L.},
	month = mar,
	year = {2016},
	note = {Publisher: The American Astronomical Society},
	pages = {89},
}

@article{yurchenko_exomol_2016,
	title = {{ExoMol} molecular line lists – {XIII}. {The} spectrum of {CaO}},
	volume = {456},
	issn = {0035-8711, 1365-2966},
	url = {https://academic.oup.com/mnras/article-lookup/doi/10.1093/mnras/stv2858},
	doi = {10.1093/mnras/stv2858},
	language = {en},
	number = {4},
	urldate = {2024-10-17},
	journal = {Monthly Notices of the Royal Astronomical Society},
	author = {Yurchenko, Sergei N. and Blissett, Audra and Asari, Usama and Vasilios, Marcus and Hill, Christian and Tennyson, Jonathan},
	month = mar,
	year = {2016},
	pages = {4524--4532},
}

@article{molliere_retrieving_2020,
	title = {Retrieving scattering clouds and disequilibrium chemistry in the atmosphere of {HR} 8799e},
	volume = {640},
	copyright = {© P. Mollière et al. 2020},
	issn = {0004-6361, 1432-0746},
	url = {https://www.aanda.org/articles/aa/abs/2020/08/aa38325-20/aa38325-20.html},
	doi = {10.1051/0004-6361/202038325},
	language = {en},
	urldate = {2024-10-17},
	journal = {Astronomy \& Astrophysics},
	author = {Mollière, P. and Stolker, T. and Lacour, S. and Otten, G. P. P. L. and Shangguan, J. and Charnay, B. and Molyarova, T. and Nowak, M. and Henning, Th and Marleau, G.-D. and Semenov, D. A. and Dishoeck, E. van and Eisenhauer, F. and Garcia, P. and Lopez, R. Garcia and Girard, J. H. and Greenbaum, A. Z. and Hinkley, S. and Kervella, P. and Kreidberg, L. and Maire, A.-L. and Nasedkin, E. and Pueyo, L. and Snellen, I. a. G. and Vigan, A. and Wang, J. and Zeeuw, P. T. de and Zurlo, A.},
	month = aug,
	year = {2020},
	note = {Publisher: EDP Sciences},
	pages = {A131},
}

@book{gray_observation_2008,
	title = {The {Observation} and {Analysis} of {Stellar} {Photospheres}},
	url = {https://ui.adsabs.harvard.edu/abs/2008oasp.book.....G},
	urldate = {2024-10-17},
	publisher = {Cambridge University Press},
	author = {Gray, David F.},
	month = jun,
	year = {2008},
	note = {Publication Title: The Observation and Analysis of Stellar Photospheres
ADS Bibcode: 2008oasp.book.....G},
}

@article{mckemmish_exomol_2019,
	title = {{ExoMol} molecular line lists – {XXXIII}. {The} spectrum of {Titanium} {Oxide}},
	volume = {488},
	issn = {0035-8711},
	url = {https://doi.org/10.1093/mnras/stz1818},
	doi = {10.1093/mnras/stz1818},
	number = {2},
	urldate = {2024-10-17},
	journal = {Monthly Notices of the Royal Astronomical Society},
	author = {McKemmish, Laura K and Masseron, Thomas and Hoeijmakers, H Jens and Pérez-Mesa, Víctor and Grimm, Simon L and Yurchenko, Sergei N and Tennyson, Jonathan},
	month = sep,
	year = {2019},
	pages = {2836--2854},
}

@article{owens_exomol_2020,
	title = {{ExoMol} line lists – {XXXVIII}. {High}-temperature molecular line list of silicon dioxide ({SiO2})},
	volume = {495},
	issn = {0035-8711},
	url = {https://doi.org/10.1093/mnras/staa1287},
	doi = {10.1093/mnras/staa1287},
	number = {2},
	urldate = {2024-10-17},
	journal = {Monthly Notices of the Royal Astronomical Society},
	author = {Owens, A and Conway, E K and Tennyson, J and Yurchenko, S N},
	month = jun,
	year = {2020},
	pages = {1927--1933},
}

@article{yurchenko_exomol_2022,
	title = {{ExoMol} line lists – {XLIV}. {Infrared} and ultraviolet line list for silicon monoxide ({28Si16O})},
	volume = {510},
	issn = {0035-8711},
	url = {https://doi.org/10.1093/mnras/stab3267},
	doi = {10.1093/mnras/stab3267},
	number = {1},
	urldate = {2024-10-17},
	journal = {Monthly Notices of the Royal Astronomical Society},
	author = {Yurchenko, Sergei N and Tennyson, Jonathan and Syme, Anna-Maree and Adam, Ahmad Y and Clark, Victoria H J and Cooper, Bridgette and Dobney, C Pria and Donnelly, Shaun T E and Gorman, Maire N and Lynas-Gray, Anthony E and Meltzer, Thomas and Owens, Alec and Qu, Qianwei and Semenov, Mikhail and Somogyi, Wilfrid and Upadhyay, Apoorva and Wright, Samuel and Zapata Trujillo, Juan C},
	month = feb,
	year = {2022},
	pages = {903--919},
}

@article{li_exomol_2019,
	title = {{ExoMol} line lists – {XXXII}. {The} rovibronic spectrum of {MgO}},
	volume = {486},
	issn = {0035-8711},
	url = {https://doi.org/10.1093/mnras/stz912},
	doi = {10.1093/mnras/stz912},
	number = {2},
	urldate = {2024-10-17},
	journal = {Monthly Notices of the Royal Astronomical Society},
	author = {Li, Heng Ying and Tennyson, Jonathan and Yurchenko, Sergei N},
	month = jun,
	year = {2019},
	pages = {2351--2365},
}

@article{patrascu_exomol_2015,
	title = {{ExoMol} molecular line lists – {IX}. {The} spectrum of {AlO}},
	volume = {449},
	issn = {0035-8711},
	url = {https://doi.org/10.1093/mnras/stv507},
	doi = {10.1093/mnras/stv507},
	number = {4},
	urldate = {2024-10-17},
	journal = {Monthly Notices of the Royal Astronomical Society},
	author = {Patrascu, Andrei T. and Yurchenko, Sergei N. and Tennyson, Jonathan},
	month = jun,
	year = {2015},
	pages = {3613--3619},
}

@article{chubb_exomolop_2021,
	title = {The {ExoMolOP} database: {Cross} sections and k-tables for molecules of interest in high-temperature exoplanet atmospheres},
	volume = {646},
	copyright = {© ESO 2021},
	issn = {0004-6361, 1432-0746},
	shorttitle = {The {ExoMolOP} database},
	url = {https://www.aanda.org/articles/aa/abs/2021/02/aa38350-20/aa38350-20.html},
	doi = {10.1051/0004-6361/202038350},
	language = {en},
	urldate = {2024-10-16},
	journal = {Astronomy \& Astrophysics},
	author = {Chubb, Katy L. and Rocchetto, Marco and Yurchenko, Sergei N. and Min, Michiel and Waldmann, Ingo and Barstow, Joanna K. and Mollière, Paul and Al-Refaie, Ahmed F. and Phillips, Mark W. and Tennyson, Jonathan},
	month = feb,
	year = {2021},
	note = {Publisher: EDP Sciences},
	pages = {A21},
}

@article{husser_new_2013,
	title = {A new extensive library of {PHOENIX} stellar atmospheres and synthetic spectra},
	volume = {553},
	copyright = {© ESO, 2013},
	issn = {0004-6361, 1432-0746},
	url = {https://www.aanda.org/articles/aa/abs/2013/05/aa19058-12/aa19058-12.html},
	doi = {10.1051/0004-6361/201219058},
	language = {en},
	urldate = {2024-10-16},
	journal = {Astronomy \& Astrophysics},
	author = {Husser, T.-O. and Berg, S. Wende-von and Dreizler, S. and Homeier, D. and Reiners, A. and Barman, T. and Hauschildt, P. H.},
	month = may,
	year = {2013},
	note = {Publisher: EDP Sciences},
	pages = {A6},
}

@article{ryabchikova_major_2015,
	title = {A major upgrade of the {VALD} database},
	volume = {90},
	issn = {1402-4896},
	url = {https://dx.doi.org/10.1088/0031-8949/90/5/054005},
	doi = {10.1088/0031-8949/90/5/054005},
	language = {en},
	number = {5},
	urldate = {2024-10-16},
	journal = {Physica Scripta},
	author = {Ryabchikova, T. and Piskunov, N. and Kurucz, R. L. and Stempels, H. C. and Heiter, U. and Pakhomov, Yu and Barklem, P. S.},
	month = apr,
	year = {2015},
	note = {Publisher: IOP Publishing},
	pages = {054005},
}

@article{morton_false_2016,
	title = {{FALSE} {POSITIVE} {PROBABILITIES} {FOR} {ALL} {KEPLER} {OBJECTS} {OF} {INTEREST}: 1284 {NEWLY} {VALIDATED} {PLANETS} {AND} 428 {LIKELY} {FALSE} {POSITIVES}},
	volume = {822},
	issn = {0004-637X},
	shorttitle = {{FALSE} {POSITIVE} {PROBABILITIES} {FOR} {ALL} {KEPLER} {OBJECTS} {OF} {INTEREST}},
	url = {https://dx.doi.org/10.3847/0004-637X/822/2/86},
	doi = {10.3847/0004-637X/822/2/86},
	language = {en},
	number = {2},
	urldate = {2024-10-16},
	journal = {The Astrophysical Journal},
	author = {Morton, Timothy D. and Bryson, Stephen T. and Coughlin, Jeffrey L. and Rowe, Jason F. and Ravichandran, Ganesh and Petigura, Erik A. and Haas, Michael R. and Batalha, Natalie M.},
	month = may,
	year = {2016},
	note = {Publisher: The American Astronomical Society},
	pages = {86},
}

@article{charpinet_compact_2011,
	title = {A compact system of small planets around a former red-giant star},
	volume = {480},
	copyright = {2011 Springer Nature Limited},
	issn = {1476-4687},
	url = {https://www.nature.com/articles/nature10631},
	doi = {10.1038/nature10631},
	language = {en},
	number = {7378},
	urldate = {2024-10-16},
	journal = {Nature},
	author = {Charpinet, S. and Fontaine, G. and Brassard, P. and Green, E. M. and Van Grootel, V. and Randall, S. K. and Silvotti, R. and Baran, A. S. and Østensen, R. H. and Kawaler, S. D. and Telting, J. H.},
	month = dec,
	year = {2011},
	note = {Publisher: Nature Publishing Group},
	pages = {496--499},
}

@book{kippenhahn_stellar_2012,
	address = {Berlin Heidelberg},
	edition = {2nd ed. 2012 edition},
	title = {Stellar {Structure} and {Evolution}},
	isbn = {978-3-642-30255-8},
	language = {English},
	publisher = {Springer},
	author = {Kippenhahn, Rudolf and Weigert, Alfred and Weiss, Achim},
	month = oct,
	year = {2012},
}

@article{elkins-tanton_lunar_2011,
	title = {The lunar magma ocean: {Reconciling} the solidification process with lunar petrology and geochronology},
	volume = {304},
	issn = {0012-821X},
	shorttitle = {The lunar magma ocean},
	url = {https://www.sciencedirect.com/science/article/pii/S0012821X11000756},
	doi = {10.1016/j.epsl.2011.02.004},
	number = {3},
	urldate = {2024-10-14},
	journal = {Earth and Planetary Science Letters},
	author = {Elkins-Tanton, Linda T. and Burgess, Seth and Yin, Qing-Zhu},
	month = apr,
	year = {2011},
	pages = {326--336},
}

@article{steenstra_possible_2020,
	title = {A possible high-temperature origin of the {Moon} and its geochemical consequences},
	volume = {538},
	issn = {0012-821X},
	url = {https://www.sciencedirect.com/science/article/pii/S0012821X20301655},
	doi = {10.1016/j.epsl.2020.116222},
	urldate = {2024-10-14},
	journal = {Earth and Planetary Science Letters},
	author = {Steenstra, E. S. and Berndt, J. and Klemme, S. and Fei, Y. and van Westrenen, W.},
	month = may,
	year = {2020},
	pages = {116222},
}

@article{wolf_vaporock_2023,
	title = {{VapoRock}: {Thermodynamics} of {Vaporized} {Silicate} {Melts} for {Modeling} {Volcanic} {Outgassing} and {Magma} {Ocean} {Atmospheres}},
	volume = {947},
	issn = {0004-637X},
	shorttitle = {{VapoRock}},
	url = {https://dx.doi.org/10.3847/1538-4357/acbcc7},
	doi = {10.3847/1538-4357/acbcc7},
	language = {en},
	number = {2},
	urldate = {2024-09-19},
	journal = {The Astrophysical Journal},
	author = {Wolf, Aaron S. and Jäggi, Noah and Sossi, Paolo A. and Bower, Dan J.},
	month = apr,
	year = {2023},
	note = {Publisher: The American Astronomical Society},
	pages = {64},
}

@article{payacan_differentiation_2023,
	title = {Differentiation of an upper crustal magma reservoir via crystal-melt separation recorded in the {San} {Gabriel} pluton, central {Chile}},
	volume = {19},
	issn = {1553-040X},
	url = {https://doi.org/10.1130/GES02535.1},
	doi = {10.1130/GES02535.1},
	number = {2},
	urldate = {2024-09-18},
	journal = {Geosphere},
	author = {Payacán, I. and Gutiérrez, F. and Bachmann, O. and Parada, M.Á.},
	month = feb,
	year = {2023},
	pages = {348--369},
}

@article{rogers_framework_2010,
	title = {A {FRAMEWORK} {FOR} {QUANTIFYING} {THE} {DEGENERACIES} {OF} {EXOPLANET} {INTERIOR} {COMPOSITIONS}},
	volume = {712},
	issn = {0004-637X},
	url = {https://dx.doi.org/10.1088/0004-637X/712/2/974},
	doi = {10.1088/0004-637X/712/2/974},
	language = {en},
	number = {2},
	urldate = {2024-09-18},
	journal = {The Astrophysical Journal},
	author = {Rogers, L. A. and Seager, S.},
	month = mar,
	year = {2010},
	note = {Publisher: The American Astronomical Society},
	pages = {974},
}

@article{swift_massradius_2011,
	title = {{MASS}–{RADIUS} {RELATIONSHIPS} {FOR} {EXOPLANETS}},
	volume = {744},
	issn = {0004-637X},
	url = {https://dx.doi.org/10.1088/0004-637X/744/1/59},
	doi = {10.1088/0004-637X/744/1/59},
	language = {en},
	number = {1},
	urldate = {2024-09-18},
	journal = {The Astrophysical Journal},
	author = {Swift, D. C. and Eggert, J. H. and Hicks, D. G. and Hamel, S. and Caspersen, K. and Schwegler, E. and Collins, G. W. and Nettelmann, N. and Ackland, G. J.},
	month = dec,
	year = {2011},
	note = {Publisher: The American Astronomical Society},
	pages = {59},
}

@article{lin_evidence_2017,
	title = {Evidence for an early wet {Moon} from experimental crystallization of the lunar magma ocean},
	volume = {10},
	copyright = {2016 Springer Nature Limited},
	issn = {1752-0908},
	url = {https://www.nature.com/articles/ngeo2845},
	doi = {10.1038/ngeo2845},
	language = {en},
	number = {1},
	urldate = {2024-09-17},
	journal = {Nature Geoscience},
	author = {Lin, Yanhao and Tronche, Elodie J. and Steenstra, Edgar S. and van Westrenen, Wim},
	month = jan,
	year = {2017},
	note = {Publisher: Nature Publishing Group},
	pages = {14--18},
}

@article{hinkel_starplanet_2018,
	title = {The {Star}–{Planet} {Connection}. {I}. {Using} {Stellar} {Composition} to {Observationally} {Constrain} {Planetary} {Mineralogy} for the 10 {Closest} {Stars}*},
	volume = {853},
	issn = {0004-637X},
	url = {https://dx.doi.org/10.3847/1538-4357/aaa5b4},
	doi = {10.3847/1538-4357/aaa5b4},
	language = {en},
	number = {1},
	urldate = {2024-09-17},
	journal = {The Astrophysical Journal},
	author = {Hinkel, Natalie R. and Unterborn, Cayman T.},
	month = jan,
	year = {2018},
	note = {Publisher: The American Astronomical Society},
	pages = {83},
}

@article{gandhi_new_2019,
	title = {New avenues for thermal inversions in atmospheres of hot {Jupiters}},
	volume = {485},
	issn = {0035-8711},
	url = {https://doi.org/10.1093/mnras/stz751},
	doi = {10.1093/mnras/stz751},
	number = {4},
	urldate = {2024-09-17},
	journal = {Monthly Notices of the Royal Astronomical Society},
	author = {Gandhi, Siddharth and Madhusudhan, Nikku},
	month = jun,
	year = {2019},
	pages = {5817--5830},
}

@article{zieba_no_2023,
	title = {No thick carbon dioxide atmosphere on the rocky exoplanet {TRAPPIST}-1 c},
	volume = {620},
	copyright = {2023 The Author(s)},
	issn = {1476-4687},
	url = {https://www.nature.com/articles/s41586-023-06232-z},
	doi = {10.1038/s41586-023-06232-z},
	language = {en},
	number = {7975},
	urldate = {2024-08-27},
	journal = {Nature},
	author = {Zieba, Sebastian and Kreidberg, Laura and Ducrot, Elsa and Gillon, Michaël and Morley, Caroline and Schaefer, Laura and Tamburo, Patrick and Koll, Daniel D. B. and Lyu, Xintong and Acuña, Lorena and Agol, Eric and Iyer, Aishwarya R. and Hu, Renyu and Lincowski, Andrew P. and Meadows, Victoria S. and Selsis, Franck and Bolmont, Emeline and Mandell, Avi M. and Suissa, Gabrielle},
	month = aug,
	year = {2023},
	note = {Publisher: Nature Publishing Group},
	pages = {746--749},
}

@misc{guimond_stars_2024,
	title = {From stars to diverse mantles, melts, crusts and atmospheres of rocky exoplanets},
	url = {http://arxiv.org/abs/2404.15427},
	doi = {10.48550/arXiv.2404.15427},
	urldate = {2024-08-27},
	publisher = {arXiv},
	author = {Guimond, Claire Marie and Wang, Haiyang and Seidler, Fabian and Sossi, Paolo and Mahajan, Aprajit and Shorttle, Oliver},
	month = apr,
	year = {2024},
	note = {arXiv:2404.15427 [astro-ph, physics:physics]},
}

@article{gale_mean_2013,
	title = {The mean composition of ocean ridge basalts},
	volume = {14},
	copyright = {©2013. American Geophysical Union. All Rights Reserved.},
	issn = {1525-2027},
	url = {https://onlinelibrary.wiley.com/doi/abs/10.1029/2012GC004334},
	doi = {10.1029/2012GC004334},
	language = {en},
	number = {3},
	urldate = {2024-08-27},
	journal = {Geochemistry, Geophysics, Geosystems},
	author = {Gale, Allison and Dalton, Colleen A. and Langmuir, Charles H. and Su, Yongjun and Schilling, Jean-Guy},
	year = {2013},
	note = {\_eprint: https://agupubs.onlinelibrary.wiley.com/doi/pdf/10.1029/2012GC004334},
	pages = {489--518},
}

@article{hans_wedepohl_composition_1995,
	title = {The composition of the continental crust},
	volume = {59},
	issn = {0016-7037},
	url = {https://www.sciencedirect.com/science/article/pii/0016703795000382},
	doi = {10.1016/0016-7037(95)00038-2},
	number = {7},
	urldate = {2024-08-27},
	journal = {Geochimica et Cosmochimica Acta},
	author = {Hans Wedepohl, K.},
	month = apr,
	year = {1995},
	pages = {1217--1232},
}

@article{boukare_deep_2022,
	title = {Deep two-phase, hemispherical magma oceans on lava planets},
	volume = {936},
	issn = {0004-637X, 1538-4357},
	url = {http://arxiv.org/abs/2205.02864},
	doi = {10.3847/1538-4357/ac8792},
	number = {2},
	urldate = {2022-09-28},
	journal = {The Astrophysical Journal},
	author = {Boukaré, C. E. and Cowan, Nicolas B. and Badro, James},
	month = sep,
	year = {2022},
	note = {arXiv:2205.02864 [astro-ph, physics:physics]},
	pages = {148},
}

@article{kurucz_atomic_1992,
	title = {Atomic and {Molecular} {Data} for {Opacity} {Calculations}},
	volume = {23},
	issn = {0185-1101},
	url = {https://ui.adsabs.harvard.edu/abs/1992RMxAA..23...45K},
	urldate = {2024-08-06},
	journal = {Revista Mexicana de Astronomia y Astrofisica, vol. 23},
	author = {Kurucz, R. L.},
	month = mar,
	year = {1992},
	note = {ADS Bibcode: 1992RMxAA..23...45K},
	pages = {45},
}

@article{grimm_helios-k_2021,
	title = {{HELIOS}-{K} 2.0 {Opacity} {Calculator} and {Open}-source {Opacity} {Database} for {Exoplanetary} {Atmospheres}},
	volume = {253},
	issn = {0067-0049},
	url = {https://dx.doi.org/10.3847/1538-4365/abd773},
	doi = {10.3847/1538-4365/abd773},
	language = {en},
	number = {1},
	urldate = {2024-08-06},
	journal = {The Astrophysical Journal Supplement Series},
	author = {Grimm, Simon L. and Malik, Matej and Kitzmann, Daniel and Guzmán-Mesa, Andrea and Hoeijmakers, H. Jens and Fisher, Chloe and Mendonça, João M. and Yurchenko, Sergey N. and Tennyson, Jonathan and Alesina, Fabien and Buchschacher, Nicolas and Burnier, Julien and Segransan, Damien and Kurucz, Robert L. and Heng, Kevin},
	month = mar,
	year = {2021},
	note = {Publisher: The American Astronomical Society},
	pages = {30},
}

@article{grimm_helios-k_2015,
	title = {{HELIOS}-{K}: {An} {Ultrafast}, {Open}-source {Opacity} {Calculator} for {Radiative} {Transfer}},
	volume = {808},
	issn = {1538-4357},
	shorttitle = {{HELIOS}-{K}},
	url = {http://arxiv.org/abs/1503.03806},
	doi = {10.1088/0004-637X/808/2/182},
	number = {2},
	urldate = {2024-08-06},
	journal = {The Astrophysical Journal},
	author = {Grimm, Simon L. and Heng, Kevin},
	month = aug,
	year = {2015},
	note = {arXiv:1503.03806 [astro-ph, physics:physics]},
	pages = {182},
}

@article{whittaker_detectability_2022,
	title = {The {Detectability} of {Rocky} {Planet} {Surface} and {Atmosphere} {Composition} with {JWST}: {The} {Case} of {LHS} 3844b},
	volume = {164},
	issn = {0004-6256, 1538-3881},
	shorttitle = {The {Detectability} of {Rocky} {Planet} {Surface} and {Atmosphere} {Composition} with {JWST}},
	url = {http://arxiv.org/abs/2207.08889},
	doi = {10.3847/1538-3881/ac9ab3},
	number = {6},
	urldate = {2024-08-06},
	journal = {The Astronomical Journal},
	author = {Whittaker, Emily A. and Malik, Matej and Ih, Jegug and Kempton, Eliza M.-R. and Mansfield, Megan and Bean, Jacob L. and Kite, Edwin S. and Koll, Daniel D. B. and Cronin, Timothy W. and Hu, Renyu},
	month = dec,
	year = {2022},
	note = {arXiv:2207.08889 [astro-ph]},
	pages = {258},
}

@article{malik_analyzing_2019,
	title = {Analyzing {Atmospheric} {Temperature} {Profiles} and {Spectra} of {M} {Dwarf} {Rocky} {Planets}},
	volume = {886},
	issn = {0004-637X},
	url = {https://ui.adsabs.harvard.edu/abs/2019ApJ...886..142M},
	doi = {10.3847/1538-4357/ab4a05},
	urldate = {2024-08-06},
	journal = {The Astrophysical Journal},
	author = {Malik, Matej and Kempton, Eliza M. -R. and Koll, Daniel D. B. and Mansfield, Megan and Bean, Jacob L. and Kite, Edwin},
	month = dec,
	year = {2019},
	note = {Publisher: IOP
ADS Bibcode: 2019ApJ...886..142M},
	pages = {142},
}

@article{hu_secondary_2024,
	title = {A secondary atmosphere on the rocky {Exoplanet} 55 {Cancri} e},
	copyright = {2024 The Author(s), under exclusive licence to Springer Nature Limited},
	issn = {1476-4687},
	url = {https://www.nature.com/articles/s41586-024-07432-x},
	doi = {10.1038/s41586-024-07432-x},
	language = {en},
	urldate = {2024-05-14},
	journal = {Nature},
	author = {Hu, Renyu and Bello-Arufe, Aaron and Zhang, Michael and Paragas, Kimberly and Zilinskas, Mantas and van Buchem, Christiaan and Bess, Michael and Patel, Jayshil and Ito, Yuichi and Damiano, Mario and Scheucher, Markus and Oza, Apurva V. and Knutson, Heather A. and Miguel, Yamila and Dragomir, Diana and Brandeker, Alexis and Demory, Brice-Olivier},
	month = may,
	year = {2024},
	note = {Publisher: Nature Publishing Group},
	pages = {1--2},
}

@article{stock_fastchem_2022,
	title = {{FastChem} 2: {An} improved computer program to determine the gas-phase chemical equilibrium composition for arbitrary element distributions},
	volume = {517},
	issn = {0035-8711, 1365-2966},
	shorttitle = {{FastChem} 2},
	url = {http://arxiv.org/abs/2206.08247},
	doi = {10.1093/mnras/stac2623},
	number = {3},
	urldate = {2024-03-20},
	journal = {Monthly Notices of the Royal Astronomical Society},
	author = {Stock, Joachim W. and Kitzmann, Daniel and Patzer, A. Beate C.},
	month = nov,
	year = {2022},
	note = {arXiv:2206.08247 [astro-ph, physics:physics]},
	pages = {4070--4080},
}

@article{leger_transiting_2009,
	title = {Transiting exoplanets from the {CoRoT} space mission: {VIII}. {CoRoT}-7b: the first super-{Earth} with measured radius},
	volume = {506},
	issn = {0004-6361, 1432-0746},
	shorttitle = {Transiting exoplanets from the {CoRoT} space mission},
	url = {http://www.aanda.org/10.1051/0004-6361/200911933},
	doi = {10.1051/0004-6361/200911933},
	language = {en},
	number = {1},
	urldate = {2024-03-13},
	journal = {Astronomy \& Astrophysics},
	author = {Léger, A. and Rouan, D. and Schneider, J. and Barge, P. and Fridlund, M. and Samuel, B. and Ollivier, M. and Guenther, E. and Deleuil, M. and Deeg, H. J. and Auvergne, M. and Alonso, R. and Aigrain, S. and Alapini, A. and Almenara, J. M. and Baglin, A. and Barbieri, M. and Bruntt, H. and Bordé, P. and Bouchy, F. and Cabrera, J. and Catala, C. and Carone, L. and Carpano, S. and Csizmadia, Sz. and Dvorak, R. and Erikson, A. and Ferraz-Mello, S. and Foing, B. and Fressin, F. and Gandolfi, D. and Gillon, M. and Gondoin, Ph. and Grasset, O. and Guillot, T. and Hatzes, A. and Hébrard, G. and Jorda, L. and Lammer, H. and Llebaria, A. and Loeillet, B. and Mayor, M. and Mazeh, T. and Moutou, C. and Pätzold, M. and Pont, F. and Queloz, D. and Rauer, H. and Renner, S. and Samadi, R. and Shporer, A. and Sotin, Ch. and Tingley, B. and Wuchterl, G. and Adda, M. and Agogu, P. and Appourchaux, T. and Ballans, H. and Baron, P. and Beaufort, T. and Bellenger, R. and Berlin, R. and Bernardi, P. and Blouin, D. and Baudin, F. and Bodin, P. and Boisnard, L. and Boit, L. and Bonneau, F. and Borzeix, S. and Briet, R. and Buey, J.-T. and Butler, B. and Cailleau, D. and Cautain, R. and Chabaud, P.-Y. and Chaintreuil, S. and Chiavassa, F. and Costes, V. and Cuna Parrho, V. and De Oliveira Fialho, F. and Decaudin, M. and Defise, J.-M. and Djalal, S. and Epstein, G. and Exil, G.-E. and Fauré, C. and Fenouillet, T. and Gaboriaud, A. and Gallic, A. and Gamet, P. and Gavalda, P. and Grolleau, E. and Gruneisen, R. and Gueguen, L. and Guis, V. and Guivarc'h, V. and Guterman, P. and Hallouard, D. and Hasiba, J. and Heuripeau, F. and Huntzinger, G. and Hustaix, H. and Imad, C. and Imbert, C. and Johlander, B. and Jouret, M. and Journoud, P. and Karioty, F. and Kerjean, L. and Lafaille, V. and Lafond, L. and Lam-Trong, T. and Landiech, P. and Lapeyrere, V. and Larqué, T. and Laudet, P. and Lautier, N. and Lecann, H. and Lefevre, L. and Leruyet, B. and Levacher, P. and Magnan, A. and Mazy, E. and Mertens, F. and Mesnager, J.-M. and Meunier, J.-C. and Michel, J.-P. and Monjoin, W. and Naudet, D. and Nguyen-Kim, K. and Orcesi, J.-L. and Ottacher, H. and Perez, R. and Peter, G. and Plasson, P. and Plesseria, J.-Y. and Pontet, B. and Pradines, A. and Quentin, C. and Reynaud, J.-L. and Rolland, G. and Rollenhagen, F. and Romagnan, R. and Russ, N. and Schmidt, R. and Schwartz, N. and Sebbag, I. and Sedes, G. and Smit, H. and Steller, M. B. and Sunter, W. and Surace, C. and Tello, M. and Tiphène, D. and Toulouse, P. and Ulmer, B. and Vandermarcq, O. and Vergnault, E. and Vuillemin, A. and Zanatta, P.},
	month = oct,
	year = {2009},
	pages = {287--302},
}

@article{batalha_pandexo_2017,
	title = {{PandExo}: {A} {Community} {Tool} for {Transiting} {Exoplanet} {Science} with {JWST} \& {HST}},
	volume = {129},
	issn = {1538-3873},
	shorttitle = {{PandExo}},
	url = {https://dx.doi.org/10.1088/1538-3873/aa65b0},
	doi = {10.1088/1538-3873/aa65b0},
	language = {en},
	number = {976},
	urldate = {2023-10-13},
	journal = {Publications of the Astronomical Society of the Pacific},
	author = {Batalha, Natasha E. and Mandell, Avi and Pontoppidan, Klaus and Stevenson, Kevin B. and Lewis, Nikole K. and Kalirai, Jason and Earl, Nick and Greene, Thomas and Albert, Loïc and Nielsen, Louise D.},
	month = apr,
	year = {2017},
	note = {Publisher: The Astronomical Society of the Pacific},
	pages = {064501},
}

@article{molliere_petitradtrans_2019,
	title = {{petitRADTRANS}: a {Python} radiative transfer package for exoplanet characterization and retrieval},
	volume = {627},
	issn = {0004-6361, 1432-0746},
	shorttitle = {{petitRADTRANS}},
	url = {http://arxiv.org/abs/1904.11504},
	doi = {10.1051/0004-6361/201935470},
	urldate = {2023-10-13},
	journal = {Astronomy \& Astrophysics},
	author = {Mollière, P. and Wardenier, J. P. and van Boekel, R. and Henning, Th and Molaverdikhani, K. and Snellen, I. A. G.},
	month = jul,
	year = {2019},
	note = {arXiv:1904.11504 [astro-ph]},
	pages = {A67},
}

@article{van_buchem_lavatmos_2023,
	title = {{LavAtmos}: {An} open-source chemical equilibrium vaporization code for lava worlds},
	volume = {58},
	copyright = {© 2023 The Authors. Meteoritics \& Planetary Science published by Wiley Periodicals LLC on behalf of The Meteoritical Society.},
	issn = {1945-5100},
	shorttitle = {{LavAtmos}},
	url = {https://onlinelibrary.wiley.com/doi/abs/10.1111/maps.13994},
	doi = {10.1111/maps.13994},
	language = {en},
	number = {8},
	urldate = {2023-09-14},
	journal = {Meteoritics \& Planetary Science},
	author = {van Buchem, Christiaan P. A. and Miguel, Yamila and Zilinskas, Mantas and van Westrenen, Wim},
	year = {2023},
	note = {\_eprint: https://onlinelibrary.wiley.com/doi/pdf/10.1111/maps.13994},
	pages = {1149--1161},
}

@incollection{hastie_alkali_1982,
	series = {{ACS} {Symposium} {Series}},
	title = {Alkali {Vapor} {Transport} in {Coal} {Conversion} and {Combustion} {Systems}},
	volume = {179},
	isbn = {978-0-8412-0689-2},
	url = {https://doi.org/10.1021/bk-1982-0179.ch034},
	urldate = {2022-03-28},
	booktitle = {Metal {Bonding} and {Interactions} in {High} {Temperature} {Systems}},
	publisher = {AMERICAN CHEMICAL SOCIETY},
	author = {Hastie, J. W. and Plante, E. R. and Bonnel, D. W.},
	month = mar,
	year = {1982},
	doi = {10.1021/bk-1982-0179.ch034},
	note = {Section: 34},
	pages = {543--600},
}

@misc{brinkman_toi-561_2022,
	title = {{TOI}-561 b: {A} {Low} {Density} {Ultra}-{Short} {Period} "{Rocky}" {Planet} around a {Metal}-{Poor} {Star}},
	shorttitle = {{TOI}-561 b},
	url = {http://arxiv.org/abs/2210.06665},
	doi = {10.48550/arXiv.2210.06665},
	urldate = {2022-10-14},
	publisher = {arXiv},
	author = {Brinkman, Casey and Weiss, Lauren M. and Dai, Fei and Huber, Daniel and Kite, Edwin S. and Valencia, Diana and Bean, Jacob L. and Beard, Corey and Behmard, Aida and Blunt, Sarah and Brady, Madison and Fulton, Benjamin and Giacalone, Steven and Howard, Andrew W. and Isaacson, Howard and Kasper, David and Lubin, Jack and MacDougall, Mason and Murphy, Joseph M. Akana and Plotnykov, Mykhalo and Polanski, Alex S. and Rice, Malena and Seifahrt, Andreas and Stefansson, Gudmundur and Sturmer, Julian},
	month = oct,
	year = {2022},
	note = {arXiv:2210.06665 [astro-ph]},
}

@incollection{palme_cosmochemical_2003,
	title = {Cosmochemical {Estimates} of {Mantle} {Composition}},
	volume = {2},
	isbn = {978-0-08-043751-4},
	language = {en},
	booktitle = {Treatise on {Geochemistry}},
	publisher = {Elsevier Ltd.},
	author = {Palme, H and O'Neill, H. St. C.},
	year = {2003},
	pages = {38},
}

@article{zilinskas_observability_2022,
	title = {Observability of evaporating lava worlds},
	volume = {661},
	issn = {0004-6361, 1432-0746},
	url = {https://www.aanda.org/10.1051/0004-6361/202142984},
	doi = {10.1051/0004-6361/202142984},
	language = {en},
	urldate = {2022-06-13},
	journal = {Astronomy \& Astrophysics},
	author = {Zilinskas, M. and van Buchem, C. P. A. and Miguel, Y. and Louca, A. and Lupu, R. and Zieba, S. and van Westrenen, W.},
	month = may,
	year = {2022},
	pages = {A126},
}

@article{stock_fastchem_2018,
	title = {{FastChem}: {A} computer program for efficient complex chemical equilibrium calculations in the neutral/ionized gas phase with applications to stellar and planetary atmospheres},
	issn = {0035-8711, 1365-2966},
	shorttitle = {{FastChem}},
	url = {https://academic.oup.com/mnras/advance-article/doi/10.1093/mnras/sty1531/5035838},
	doi = {10.1093/mnras/sty1531},
	language = {en},
	urldate = {2022-06-01},
	journal = {Monthly Notices of the Royal Astronomical Society},
	author = {Stock, Joachim W and Kitzmann, Daniel and Patzer, A Beate C and Sedlmayr, Erwin},
	month = jun,
	year = {2018},
}

@article{malik_helios_2017,
	title = {{HELIOS}: {AN} {OPEN}-{SOURCE}, {GPU}-{ACCELERATED} {RADIATIVE} {TRANSFER} {CODE} {FOR} {SELF}-{CONSISTENT} {EXOPLANETARY} {ATMOSPHERES}},
	volume = {153},
	issn = {1538-3881},
	shorttitle = {{HELIOS}},
	url = {https://doi.org/10.3847/1538-3881/153/2/56},
	doi = {10.3847/1538-3881/153/2/56},
	language = {en},
	number = {2},
	urldate = {2022-06-01},
	journal = {The Astronomical Journal},
	author = {Malik, Matej and Grosheintz, Luc and Mendonça, João M. and Grimm, Simon L. and Lavie, Baptiste and Kitzmann, Daniel and Tsai, Shang-Min and Burrows, Adam and Kreidberg, Laura and Bedell, Megan and Bean, Jacob L. and Stevenson, Kevin B. and Heng, Kevin},
	month = jan,
	year = {2017},
	note = {Publisher: American Astronomical Society},
	pages = {56},
}

@article{wang_detailed_2022,
	title = {Detailed chemical compositions of planet-hosting stars: {II}. {Exploration} of the interiors of terrestrial-type exoplanets},
	shorttitle = {Detailed chemical compositions of planet-hosting stars},
	url = {http://arxiv.org/abs/2204.09558},
	urldate = {2022-04-21},
	journal = {arXiv:2204.09558 [astro-ph]},
	author = {Wang, Haiyang S. and Quanz, Sascha P. and Yong, David and Liu, Fan and Seidler, Fabian and Acuña, Lorena and Mojzsis, Stephen J.},
	month = apr,
	year = {2022},
	note = {arXiv: 2204.09558},
}

@article{hastie_predictive_1985,
	title = {A {Predictive} {Phase} {Equilibrium} {Model} {For} {Multicomponent} {Oxide} {Mixtures} {Part} {II} {Oxides} {Of} {Na}-{K}-{Ca}-{Mg}-{Al}-{Si}},
	volume = {19},
	number = {3},
	journal = {High Temp.Sci.},
	author = {Hastie, J.W. and Bonnell, D. W.},
	year = {1985},
	pages = {275--306},
}

@article{henning_highly_2018,
	title = {Highly {Volcanic} {Exoplanets}, {Lava} {Worlds}, and {Magma} {Ocean} {Worlds}: {An} {Emerging} {Class} of {Dynamic} {Exoplanets} of {Significant} {Scientific} {Priority}},
	shorttitle = {Highly {Volcanic} {Exoplanets}, {Lava} {Worlds}, and {Magma} {Ocean} {Worlds}},
	url = {http://arxiv.org/abs/1804.05110},
	urldate = {2022-04-14},
	journal = {arXiv:1804.05110 [astro-ph, physics:physics]},
	author = {Henning, Wade G. and Renaud, Joseph P. and Saxena, Prabal and Whelley, Patrick L. and Mandell, Avi M. and Matsumura, Soko and Glaze, Lori S. and Hurford, Terry A. and Livengood, Timothy A. and Hamilton, Christopher W. and Efroimsky, Michael and Makarov, Valeri V. and Berghea, Ciprian T. and Guzewich, Scott D. and Tsigaridis, Kostas and Arney, Giada N. and Cremons, Daniel R. and Kane, Stephen R. and Bleacher, Jacob E. and Kopparapu, Ravi K. and Kohler, Erika and Lee, Yuni and Rushby, Andrew and Kuang, Weijia and Barnes, Rory and Richardson, Jacob A. and Driscoll, Peter and Schmerr, Nicholas C. and Del Genio, Anthony D. and Davies, Ashley Gerard and Kaltenegger, Lisa and Elkins-Tanton, Linda and Fujii, Yuka and Schaefer, Laura and Ranjan, Sukrit and Quintana, Elisa and Barclay, Thomas S. and Hamano, Keiko and Petro, Noah E. and Kendall, Jordan D. and Lopez, Eric D. and Sasselov, Dimitar D.},
	month = apr,
	year = {2018},
	note = {arXiv: 1804.05110},
}

@article{ghiorso_h2oco2_2015,
	title = {An {H2O}–{CO2} mixed fluid saturation model compatible with rhyolite-{MELTS}},
	volume = {169},
	issn = {1432-0967},
	url = {https://doi.org/10.1007/s00410-015-1141-8},
	doi = {10.1007/s00410-015-1141-8},
	language = {en},
	number = {6},
	urldate = {2022-04-05},
	journal = {Contributions to Mineralogy and Petrology},
	author = {Ghiorso, Mark S. and Gualda, Guilherme A. R.},
	month = jun,
	year = {2015},
	pages = {53},
}

@article{ito_theoretical_2015,
	title = {{THEORETICAL} {EMISSION} {SPECTRA} {OF} {ATMOSPHERES} {OF} {HOT} {ROCKY} {SUPER}-{EARTHS}},
	volume = {801},
	issn = {0004-637X},
	url = {https://doi.org/10.1088/0004-637x/801/2/144},
	doi = {10.1088/0004-637X/801/2/144},
	language = {en},
	number = {2},
	urldate = {2022-03-16},
	journal = {The Astrophysical Journal},
	author = {Ito, Yuichi and Ikoma, Masahiro and Kawahara, Hajime and Nagahara, Hiroko and Kawashima, Yui and Nakamoto, Taishi},
	month = mar,
	year = {2015},
	note = {Publisher: American Astronomical Society},
	pages = {144},
}

@book{chase_nist-janaf_1998,
	edition = {4},
	title = {{NIST}-{JANAF} {Thermochemical} {Tables}},
	volume = {1},
	series = {Journal of Physical and Chemical Reference Data Monograph},
	url = {http://adsabs.harvard.edu/pdf/1971LPSC....2.1367D},
	urldate = {2021-10-14},
	publisher = {American Chemical Society},
	author = {Chase, Malcolm W.},
	year = {1998},
}

@article{putirka_polluted_2021,
	title = {Polluted white dwarfs reveal exotic mantle rock types on exoplanets in our solar neighborhood},
	volume = {12},
	issn = {2041-1723},
	url = {https://www.nature.com/articles/s41467-021-26403-8},
	doi = {10.1038/s41467-021-26403-8},
	language = {en},
	number = {1},
	urldate = {2021-11-08},
	journal = {Nature Communications},
	author = {Putirka, Keith D. and Xu, Siyi},
	month = dec,
	year = {2021},
	pages = {6168},
}

@article{hastie_predictive_1986,
	title = {A predictive thermodynamic model of oxide and halide glass phase equilibria},
	volume = {84},
	issn = {00223093},
	url = {https://linkinghub.elsevier.com/retrieve/pii/0022309386907726},
	doi = {10.1016/0022-3093(86)90772-6},
	language = {en},
	number = {1-3},
	urldate = {2021-09-15},
	journal = {Journal of Non-Crystalline Solids},
	author = {Hastie, J.W. and Bonnell, D.W.},
	month = jul,
	year = {1986},
	pages = {151--158},
}

@article{gualda_rhyolite-melts_2012,
	title = {Rhyolite-{MELTS}: a {Modified} {Calibration} of {MELTS} {Optimized} for {Silica}-rich, {Fluid}-bearing {Magmatic} {Systems}},
	volume = {53},
	issn = {0022-3530},
	shorttitle = {Rhyolite-{MELTS}},
	url = {https://doi.org/10.1093/petrology/egr080},
	doi = {10.1093/petrology/egr080},
	number = {5},
	urldate = {2021-09-13},
	journal = {Journal of Petrology},
	author = {Gualda, Guilherme A. R. and Ghiorso, Mark S. and Lemons, Robin V. and Carley, Tamara L.},
	month = may,
	year = {2012},
	pages = {875--890},
}

@article{putirka_compositional_2021,
	title = {Compositional {Diversity} of {Rocky} {Exoplanets}},
	url = {http://arxiv.org/abs/2108.08383},
	urldate = {2021-08-20},
	journal = {arXiv:2108.08383 [astro-ph, physics:physics]},
	author = {Putirka, Keith and Dorn, Caroline and Hinkel, Natalie and Unterborn, Cayman},
	month = aug,
	year = {2021},
	note = {arXiv: 2108.08383},
}

@article{xu_exogeology_2021,
	title = {Exogeology from {Polluted} {White} {Dwarfs}},
	url = {http://arxiv.org/abs/2108.08384},
	urldate = {2021-08-20},
	journal = {arXiv:2108.08384 [astro-ph]},
	author = {Xu, Siyi and Bonsor, Amy},
	month = aug,
	year = {2021},
	note = {arXiv: 2108.08384},
}

@article{dorn_can_2015,
	title = {Can we constrain the interior structure of rocky exoplanets from mass and radius measurements?},
	volume = {577},
	copyright = {© ESO, 2015},
	issn = {0004-6361, 1432-0746},
	url = {http://www.aanda.org/articles/aa/abs/2015/05/aa24915-14/aa24915-14.html},
	doi = {10.1051/0004-6361/201424915},
	language = {en},
	urldate = {2021-07-06},
	journal = {Astronomy \& Astrophysics},
	author = {Dorn, Caroline and Khan, Amir and Heng, Kevin and Connolly, James A. D. and Alibert, Yann and Benz, Willy and Tackley, Paul},
	month = may,
	year = {2015},
	note = {Publisher: EDP Sciences},
	pages = {A83},
}

@article{santos_constraining_2017,
	title = {Constraining planet structure and composition from stellar chemistry: trends in different stellar populations},
	volume = {608},
	issn = {0004-6361, 1432-0746},
	shorttitle = {Constraining planet structure and composition from stellar chemistry},
	url = {http://arxiv.org/abs/1711.00777},
	doi = {10.1051/0004-6361/201731359},
	urldate = {2021-07-01},
	journal = {Astronomy \& Astrophysics},
	author = {Santos, N. C. and Adibekyan, V. and Dorn, C. and Mordasini, C. and Noack, L. and Barros, S. C. C. and Delgado-Mena, E. and Demangeon, O. and Faria, J. and Israelian, G. and Sousa, S. G.},
	month = dec,
	year = {2017},
	note = {arXiv: 1711.00777},
	pages = {A94},
}

@article{hakim_mineralogy_2019,
	title = {Mineralogy, {Structure}, and {Habitability} of {Carbon}-{Enriched} {Rocky} {Exoplanets}: {A} {Laboratory} {Approach}},
	volume = {19},
	issn = {1531-1074},
	shorttitle = {Mineralogy, {Structure}, and {Habitability} of {Carbon}-{Enriched} {Rocky} {Exoplanets}},
	url = {https://www.liebertpub.com/doi/10.1089/ast.2018.1930},
	doi = {10.1089/ast.2018.1930},
	number = {7},
	urldate = {2021-06-17},
	journal = {Astrobiology},
	author = {Hakim, Kaustubh and Spaargaren, Rob and Grewal, Damanveer S. and Rohrbach, Arno and Berndt, Jasper and Dominik, Carsten and van Westrenen, Wim},
	month = jul,
	year = {2019},
	note = {Publisher: Mary Ann Liebert, Inc., publishers},
	pages = {867--884},
}

@article{klein_rocky_2011,
	title = {{ROCKY} {EXTRASOLAR} {PLANETARY} {COMPOSITIONS} {DERIVED} {FROM} {EXTERNALLY} {POLLUTED} {WHITE} {DWARFS}},
	volume = {741},
	issn = {0004-637X},
	url = {https://doi.org/10.1088/0004-637x/741/1/64},
	doi = {10.1088/0004-637X/741/1/64},
	language = {en},
	number = {1},
	urldate = {2021-06-16},
	journal = {The Astrophysical Journal},
	author = {Klein, B. and Jura, M. and Koester, D. and Zuckerman, B.},
	month = oct,
	year = {2011},
	note = {Publisher: American Astronomical Society},
	pages = {64},
}

@article{putirka_composition_2019,
	title = {The composition and mineralogy of rocky exoplanets: {A} survey of {\textgreater}4000 stars from the {Hypatia} {Catalog}},
	volume = {104},
	issn = {1945-3027},
	shorttitle = {The composition and mineralogy of rocky exoplanets},
	url = {http://www.degruyter.com/document/doi/10.2138/am-2019-6787/html},
	doi = {10.2138/am-2019-6787},
	language = {en},
	number = {6},
	urldate = {2021-05-20},
	journal = {American Mineralogist},
	author = {Putirka, Keith D. and Rarick, John C.},
	month = jun,
	year = {2019},
	note = {Publisher: Mineralogical Society of America
Section: American Mineralogist},
	pages = {817--829},
}

@article{hinkel_hypatia_2017,
	title = {The {Hypatia} {Catalog} {Database}: {A} {Web}-{Based} {Interface} for {Exploring} {Stellar} {Abundances}},
	shorttitle = {The {Hypatia} {Catalog} {Database}},
	url = {http://arxiv.org/abs/1712.04944},
	urldate = {2021-05-12},
	journal = {arXiv:1712.04944 [astro-ph]},
	author = {Hinkel, Natalie R. and Burger, Dan},
	month = dec,
	year = {2017},
	note = {arXiv: 1712.04944},
}

@article{schaefer_chemistry_2009,
	title = {Chemistry of {Silicate} {Atmospheres} of {Evaporating} {Super}-{Earths}},
	volume = {703},
	issn = {0004-637X},
	url = {http://adsabs.harvard.edu/abs/2009ApJ...703L.113S},
	doi = {10.1088/0004-637X/703/2/L113},
	urldate = {2021-05-07},
	journal = {The Astrophysical Journal Letters},
	author = {Schaefer, Laura and Fegley, Bruce},
	month = oct,
	year = {2009},
	pages = {L113--L117},
}

@article{carter-bond_low_2012,
	title = {{LOW} {Mg}/{Si} {PLANETARY} {HOST} {STARS} {AND} {THEIR} {Mg}-{DEPLETED} {TERRESTRIAL} {PLANETS}},
	volume = {747},
	issn = {2041-8205},
	url = {https://doi.org/10.1088/2041-8205/747/1/l2},
	doi = {10.1088/2041-8205/747/1/L2},
	language = {en},
	number = {1},
	urldate = {2021-05-06},
	journal = {The Astrophysical Journal},
	author = {Carter-Bond, Jade C. and O{\textbackslash}textquotesingleBrien, David P. and Mena, Elisa Delgado and Israelian, Garik and Santos, Nuno C. and Hernández, Jonay I. González},
	month = feb,
	year = {2012},
	note = {Publisher: American Astronomical Society},
	pages = {L2},
}

@article{hakim_thermal_2019,
	title = {Thermal evolution of rocky exoplanets with a graphite outer shell},
	volume = {630},
	copyright = {© ESO 2019},
	issn = {0004-6361, 1432-0746},
	url = {http://www.aanda.org/articles/aa/abs/2019/10/aa35714-19/aa35714-19.html},
	doi = {10.1051/0004-6361/201935714},
	language = {en},
	urldate = {2021-05-05},
	journal = {Astronomy \& Astrophysics},
	author = {Hakim, Kaustubh and Berg, Arie van den and Vazan, Allona and Höning, Dennis and Westrenen, Wim van and Dominik, Carsten},
	month = oct,
	year = {2019},
	note = {Publisher: EDP Sciences},
	pages = {A152},
}

@article{miguel_compositions_2011,
	title = {{COMPOSITIONS} {OF} {HOT} {SUPER}-{EARTH} {ATMOSPHERES}: {EXPLORING} {KEPLER} {CANDIDATES}},
	volume = {742},
	issn = {2041-8205},
	shorttitle = {{COMPOSITIONS} {OF} {HOT} {SUPER}-{EARTH} {ATMOSPHERES}},
	url = {https://doi.org/10.1088/2041-8205/742/2/l19},
	doi = {10.1088/2041-8205/742/2/L19},
	language = {en},
	number = {2},
	urldate = {2021-05-04},
	journal = {The Astrophysical Journal},
	author = {Miguel, Y. and Kaltenegger, L. and Fegley, B. and Schaefer, L.},
	month = nov,
	year = {2011},
	note = {Publisher: American Astronomical Society},
	pages = {L19},
}

@article{bonsor_host-star_2021,
	title = {Host-star and exoplanet compositions: a pilot study using a wide binary with a polluted white dwarf},
	volume = {503},
	issn = {0035-8711},
	shorttitle = {Host-star and exoplanet compositions},
	url = {https://doi.org/10.1093/mnras/stab370},
	doi = {10.1093/mnras/stab370},
	number = {2},
	urldate = {2021-05-04},
	journal = {Monthly Notices of the Royal Astronomical Society},
	author = {Bonsor, Amy and Jofré, Paula and Shorttle, Oliver and Rogers, Laura K and Xu, Siyi and Melis, Carl},
	month = may,
	year = {2021},
	pages = {1877--1883},
}

@article{asimow_algorithmic_1998,
	title = {Algorithmic modifications extending {MELTS} to calculate subsolidus phase relations},
	volume = {83},
	issn = {1945-3027},
	url = {http://www.degruyter.com/document/doi/10.2138/am-1998-9-1022/html},
	doi = {10.2138/am-1998-9-1022},
	language = {en},
	number = {9-10},
	urldate = {2021-04-22},
	journal = {American Mineralogist},
	author = {Asimow, Paul D. and Ghiorso, Mark S.},
	month = sep,
	year = {1998},
	note = {Publisher: Mineralogical Society of America
Section: American Mineralogist},
	pages = {1127--1132},
}

@article{ghiorso_chemical_1985-1,
	title = {Chemical mass transfer in magmatic processes {II}. {Applications} in equilibrium crystallization, fractionation and assimilation},
	volume = {90},
	issn = {1432-0967},
	url = {https://doi.org/10.1007/BF00378255},
	doi = {10.1007/BF00378255},
	language = {en},
	number = {2},
	urldate = {2021-04-20},
	journal = {Contributions to Mineralogy and Petrology},
	author = {Ghiorso, Mark S. and Carmichael, Ian S. E.},
	month = jul,
	year = {1985},
	pages = {121--141},
}

@article{kite_atmosphere-interior_2016,
	title = {Atmosphere-interior exchange on hot rocky exoplanets},
	volume = {828},
	issn = {1538-4357},
	url = {http://arxiv.org/abs/1606.06740},
	doi = {10.3847/0004-637X/828/2/80},
	number = {2},
	urldate = {2021-04-15},
	journal = {The Astrophysical Journal},
	author = {Kite, Edwin S. and Fegley Jr., Bruce and Schaefer, Laura and Gaidos, Eric},
	month = sep,
	year = {2016},
	note = {arXiv: 1606.06740},
	pages = {80},
}

@article{nguyen_modelling_2020,
	title = {Modelling the atmosphere of lava planet {K2}-141b: implications for low- and high-resolution spectroscopy},
	volume = {499},
	issn = {0035-8711, 1365-2966},
	shorttitle = {Modelling the atmosphere of lava planet {K2}-141b},
	url = {https://academic.oup.com/mnras/article/499/4/4605/5948140},
	doi = {10.1093/mnras/staa2487},
	language = {en},
	number = {4},
	urldate = {2021-03-31},
	journal = {Monthly Notices of the Royal Astronomical Society},
	author = {Nguyen, T Giang and Cowan, Nicolas B and Banerjee, Agnibha and Moores, John E},
	month = dec,
	year = {2020},
	pages = {4605--4612},
}

@article{ghiorso_pmelts_2002,
	title = {The {pMELTS}: {A} revision of {MELTS} for improved calculation of phase relations and major element partitioning related to partial melting of the mantle to 3 {GPa}},
	volume = {3},
	copyright = {Copyright 2002 by the American Geophysical Union.},
	issn = {1525-2027},
	shorttitle = {The {pMELTS}},
	url = {https://agupubs.onlinelibrary.wiley.com/doi/abs/10.1029/2001GC000217},
	doi = {https://doi.org/10.1029/2001GC000217},
	language = {en},
	number = {5},
	urldate = {2021-03-30},
	journal = {Geochemistry, Geophysics, Geosystems},
	author = {Ghiorso, Mark S. and Hirschmann, Marc M. and Reiners, Peter W. and Kress, Victor C.},
	year = {2002},
	pages = {1--35},
}

@article{ghiorso_chemical_1995,
	title = {Chemical mass transfer in magmatic processes {IV}. {A} revised and internally consistent thermodynamic model for the interpolation and extrapolation of liquid-solid equilibria in magmatic systems at elevated temperatures and pressures},
	volume = {119},
	issn = {1432-0967},
	url = {https://doi.org/10.1007/BF00307281},
	doi = {10.1007/BF00307281},
	language = {en},
	number = {2},
	urldate = {2021-03-30},
	journal = {Contributions to Mineralogy and Petrology},
	author = {Ghiorso, Mark S. and Sack, Richard O.},
	month = mar,
	year = {1995},
	pages = {197--212},
}

@article{schaefer_thermodynamic_2004,
	title = {A thermodynamic model of high temperature lava vaporization on {Io}},
	volume = {169},
	issn = {00191035},
	url = {https://linkinghub.elsevier.com/retrieve/pii/S0019103503003191},
	doi = {10.1016/j.icarus.2003.08.023},
	language = {en},
	number = {1},
	urldate = {2021-03-25},
	journal = {Icarus},
	author = {Schaefer, Laura and Fegley Jr., Bruce},
	month = may,
	year = {2004},
	pages = {216--241},
}

@article{fegley_vaporization_1987,
	title = {A vaporization model for iron/silicate fractionation in the {Mercury} protoplanet},
	volume = {82},
	issn = {0012821X},
	url = {https://linkinghub.elsevier.com/retrieve/pii/0012821X87901968},
	doi = {10.1016/0012-821X(87)90196-8},
	language = {en},
	number = {3-4},
	urldate = {2021-03-25},
	journal = {Earth and Planetary Science Letters},
	author = {Fegley, Bruce and Cameron, A.G.W.},
	month = apr,
	year = {1987},
	pages = {207--222},
}




\appendix

\section{Melt compositions}\label{app:melt_composition}

The melt compositions are calculated by multiplying their original BSE weight percentage, as given in \citet{palme_cosmochemical_2003}, by different factors (0.1, 0.5, 2, and 10). This leads to weight percentages of greater than 100 for some species. This is not an issue for LavAtmos due to the fact that the composition is normalized internally. However, it may give a wrong impression of the actual abundance that is being tested. Hence, we provide a table below showing the varied weight percentages for each of the species for which we present results  (indicated with \textit{Before norm.}) as well as their normalized counterparts (indicated with \textit{After norm.}). Note that SiO$_2$ abundances lower than BSE could not be tested due to limitations of the MELTS code.  

\begin{table*}
\centering
\caption{\textbf{Melt composition variation}}
\label{tab:compositions}
\resizebox{\linewidth}{!}{%
\begin{tabular}{lccccccccc}
 & \multicolumn{9}{c}{\textbf{Multiplier}} \\
\multicolumn{1}{l|}{\textbf{}} & \multicolumn{2}{c|}{\textbf{x0.10}} & \multicolumn{2}{c|}{\textbf{x0.50}} & \multicolumn{1}{c|}{\textbf{x1}} & \multicolumn{2}{c|}{\textbf{x2}} & \multicolumn{2}{c|}{\textbf{x10}} \\
\multicolumn{1}{l|}{\textbf{Species}} & Before norm. [\%] & \multicolumn{1}{c|}{After norm. [\%]} & Before norm. [\%] & \multicolumn{1}{c|}{After norm. [\%]} & \multicolumn{1}{c|}{\textbf{BSE}} & Before norm. [\%] & \multicolumn{1}{c|}{After norm. [\%]} & Before norm. [\%] & \multicolumn{1}{c|}{After norm. [\%]} \\ \hline
\multicolumn{1}{l|}{} & \multicolumn{1}{l}{} & \multicolumn{1}{l|}{} & \multicolumn{1}{l}{} & \multicolumn{1}{l|}{} & \multicolumn{1}{l|}{} & \multicolumn{1}{l}{} & \multicolumn{1}{l|}{} & \multicolumn{1}{l}{} & \multicolumn{1}{l|}{} \\
\multicolumn{1}{l|}{SiO$_2$} & - & \multicolumn{1}{c|}{-} & - & \multicolumn{1}{c|}{-} & \multicolumn{1}{c|}{\textbf{4.54E+01}} & 9.08E+01 & \multicolumn{1}{c|}{6.24E+01} & 4.54E+02 & \multicolumn{1}{c|}{8.93E+01} \\
\multicolumn{1}{l|}{TiO$_2$} & 2.10E-02 & \multicolumn{1}{c|}{2.10E-02} & 1.05E-01 & \multicolumn{1}{c|}{1.05E-01} & \multicolumn{1}{c|}{\textbf{2.10E-01}} & 4.20E-01 & \multicolumn{1}{c|}{4.19E-01} & 2.10E+00 & \multicolumn{1}{c|}{2.06E+00} \\
\multicolumn{1}{l|}{Na$_2$O} & 3.49E-02 & \multicolumn{1}{c|}{3.50E-02} & 1.75E-01 & \multicolumn{1}{c|}{1.75E-01} & \multicolumn{1}{c|}{\textbf{3.49E-01}} & 6.98E-01 & \multicolumn{1}{c|}{6.96E-01} & 3.49E+00 & \multicolumn{1}{c|}{3.38E+00} \\
\multicolumn{1}{l|}{K$_2$O} & 3.10E-03 & \multicolumn{1}{c|}{3.10E-03} & 1.55E-02 & \multicolumn{1}{c|}{1.55E-02} & \multicolumn{1}{c|}{\textbf{3.10E-02}} & 6.20E-02 & \multicolumn{1}{c|}{6.20E-02} & 3.10E-01 & \multicolumn{1}{c|}{3.09E-01}
\end{tabular}%
}
\end{table*}

\section{Opacity data}

\begin{table*}
\centering
\caption{List of opacities and their sources used to calculate the temperature-pressure profiles and emission spectra in this work. Those with DACE as indicated source were taken directly from the DACE database (\url{https://dace.unige.ch/}). For the other species, the opacities were calculated using HELIOS-K (\url{https://github.com/exoclime/HELIOS-K}) \citep{grimm_helios-k_2015,grimm_helios-k_2021}.}
\label{tab:opacity_sources}
\begin{tabular}{llll}
Species                  & Source        & Line list & Line list reference          \\ \hline  \hline
                         &               &           &                              \\
Al                       & DACE          & VALD      & \citet{ryabchikova_major_2015}    \\
AlO                      & HELIOS-K      & ATP       & \citet{patrascu_exomol_2015}       \\
Ca                       & DACE          & VALD      & \citet{ryabchikova_major_2015}    \\
CaO                      & HELIOS-K      & VBATHY    & \citet{yurchenko_exomol_2016}      \\
Fe                       & DACE          & VALD      & \citet{ryabchikova_major_2015}    \\
K                        & DACE          & VALD      & \citet{ryabchikova_major_2015}    \\
Mg                       & DACE          & Kurucz    & \citet{kurucz_atomic_1992}                \\
MgO                      & HELIOS-K      & LiTY      & \citet{li_exomol_2019}             \\
Na                       & DACE          & VALD      & \citet{ryabchikova_major_2015}    \\
Si                       & DACE          & VALD      & \citet{ryabchikova_major_2015}    \\
SiO                      & HELIOS-K      & SiOUVenIR & \citet{yurchenko_exomol_2022}      \\
SiO$_2$                  & DACE          & OYT3      & \citet{owens_exomol_2020}          \\
Ti                       & DACE          & VALD      & \citet{ryabchikova_major_2015}    \\
TiO                      & HELIOS-K      & Toto      & \citet{mckemmish_exomol_2019}      \\
                         &               &           &                              \\
Scattering and continuum &               &           &                              \\ \hline
                         &               &           &                              \\
O$_2$                    & Scattering    &           & \citet{gray_observation_2008}                  \\
O$_2$-O$_2$              & petitRADTRANS &           & \citet{molliere_petitradtrans_2019,molliere_retrieving_2020}
\end{tabular}
\end{table*}

\begin{figure*}
    \centering
    \includegraphics[width=0.5\linewidth]{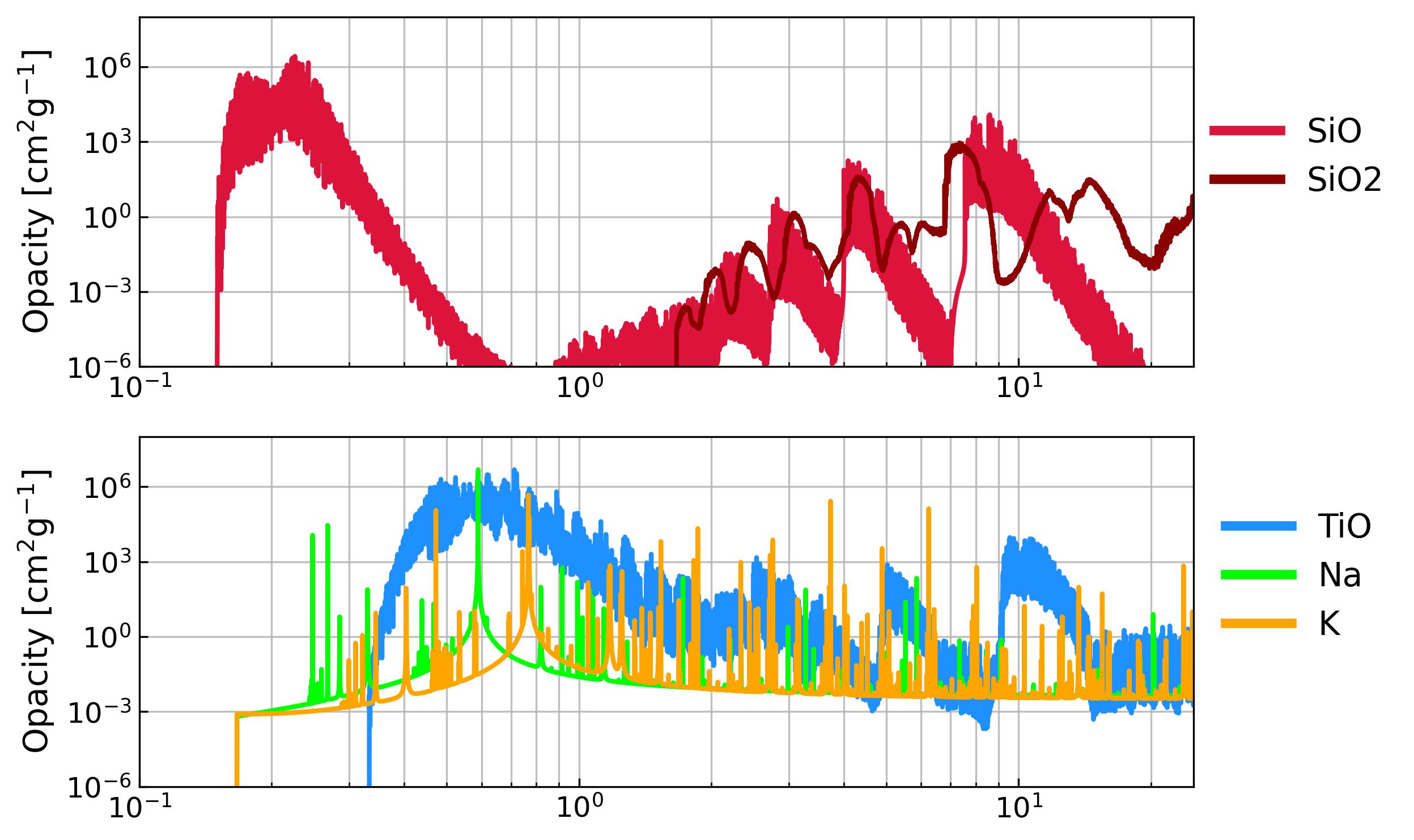}
    \caption{\textbf{Opacities of important atmospheric species:} The unweighed opacities of SiO, SiO$_2$, TiO, Na, and K are shown here for a temperature of 2700 K and 1e-2 bar. See Table \ref{tab:opacity_sources} for the line lists used to calculate the opacities shown here.}
    \label{fig:opacites}
\end{figure*}

\section{Stellar spectra}

\begin{figure*}[H]
    \centering
    \includegraphics[width=0.5\linewidth]{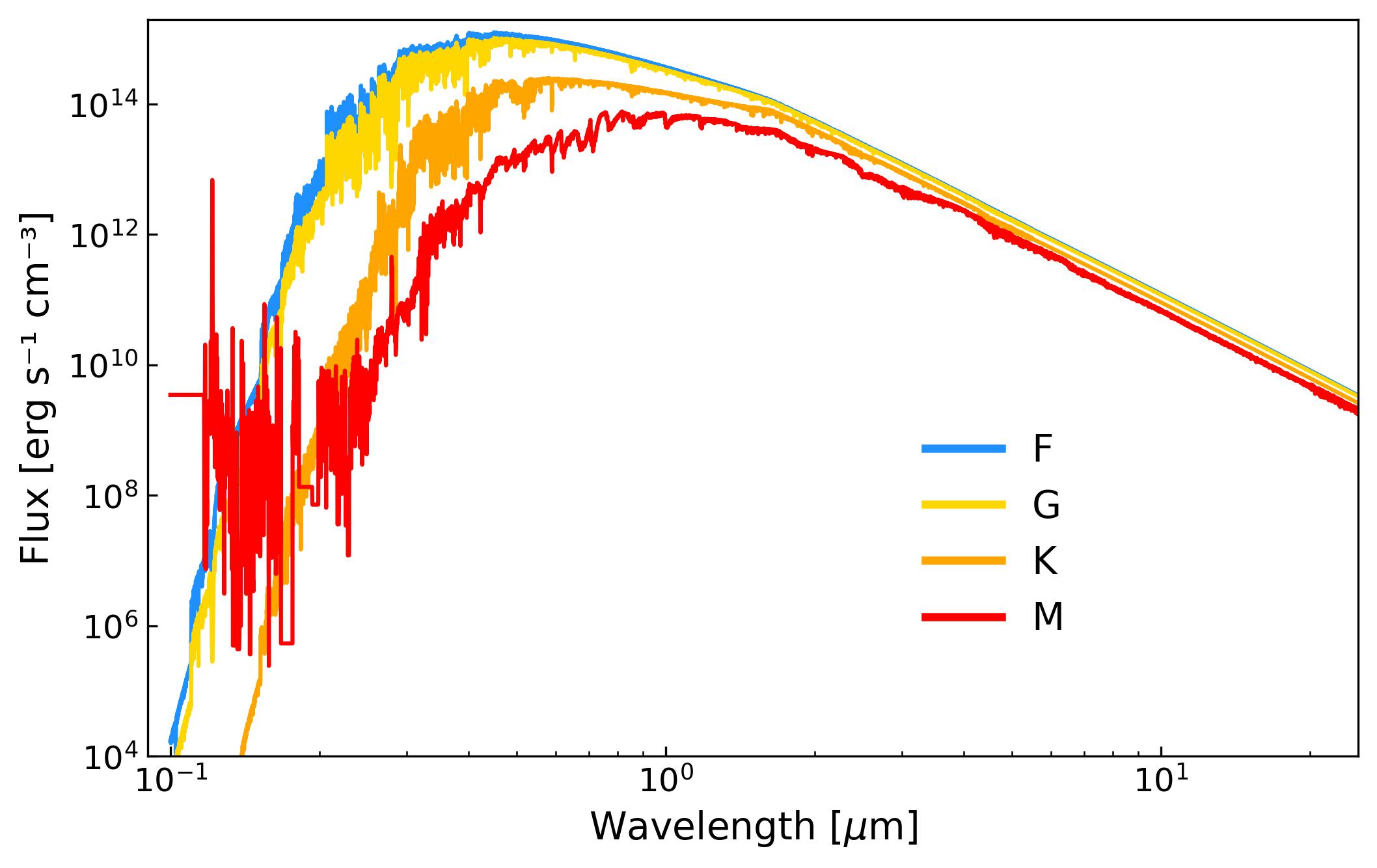}
    \caption{\textbf{Stellar spectra used for each of stellar type:} The spectra for the F, G, and K stars were calculated using the \texttt{PHOENIX} models. For the M type star we combined the \texttt{PHOENIX} model with scaled \texttt{MUSCLES} shortwave spectra.}
    \label{fig:stellar_spectra}
\end{figure*}



\section{\texorpdfstring{SiO$_2$}{SiO2} activity changing with alkali abundances}

\begin{figure*}[H]
    \centering
    \includegraphics[width=0.75\linewidth]{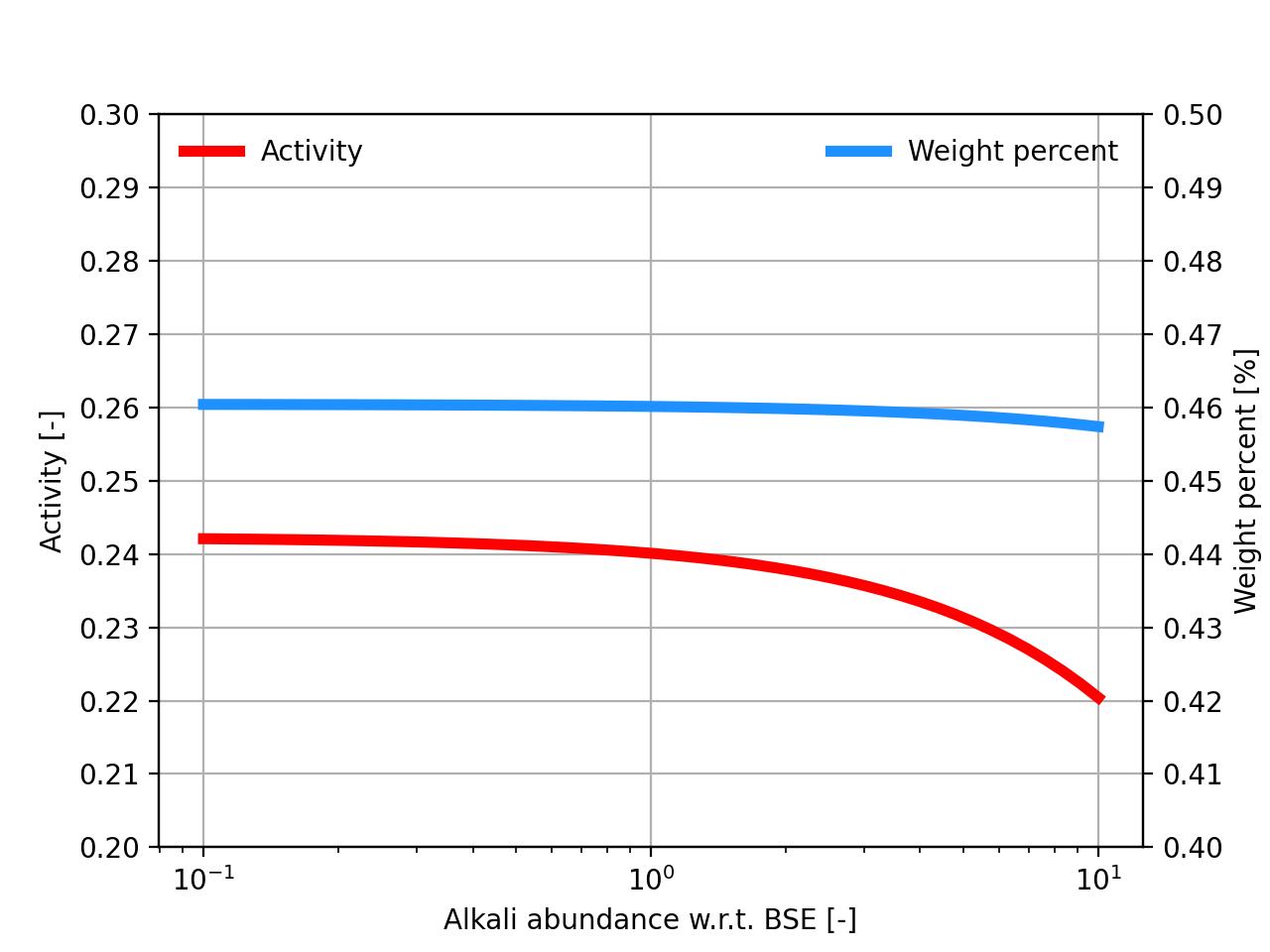}
    \caption{\textbf{SiO$_2$ melt activity as a function of changing alkali abundances:} Calculated using ThermoEngine, the python wrapper to MELTS \citep{ghiorso_chemical_1995}.}
    \label{fig:sio2_activity}
\end{figure*}


\bsp	
\label{lastpage}
\end{document}